\documentclass[twocolumn, tighten, twocolappendix]{aastex631_mod}             

\usepackage[T1]{fontenc}
\usepackage{microtype}
\usepackage{newtxtext}
\usepackage[varvw]{newtxmath}
\usepackage{mathtools}
\usepackage{amsmath}
\usepackage{bm}
\usepackage{graphicx}
\usepackage{acronym}
\usepackage{booktabs}

\newacro{DF}{distribution function}
\newacroplural{DFs}{distribution functions}
\newcommand{\DF}{\ac{DF}}

\newacro{CBE}{collisionless Boltzmann equation}

\newacro{DM}{Dark Matter}

\newacro{COM}{centre of mass}

\newcommand{\ri}{\mathrm{i}}

\newcommand{\re}{\mathrm{e}}
\newcommand{\rd}{\mathrm{d}}
\newcommand{\rD}{\mathrm{D}}

\newcommand{\rt}{\mathrm{t}}
\newcommand{\rs}{\mathrm{s}}

\newcommand{\rf}{\mathrm{f}}

\newcommand{\bT}{\bm{\theta}}
\newcommand{\bJ}{\mathbf{J}}
\newcommand{\bx}{\mathbf{x}}
\renewcommand{\bv}{\mathbf{v}}

\newcommand{\bxp}{\mathbf{x}^{\prime}}
\newcommand{\bvp}{\mathbf{v}^{\prime}}

\newcommand{\bn}{\mathbf{n}}

\newcommand{\bJp}{\mathbf{J}^{\prime}}

\newcommand{\bTp}{\bT^{\prime}}
\newcommand{\tbJ}{\widetilde{\mathbf{J}}}
\newcommand{\tbn}{\widetilde{\mathbf{n}}}
\newcommand{\bO}{\mathbf{\Omega}}
\newcommand{\tbO}{\widetilde{\mathbf{\Omega}}}

\newcommand{\bM}{\mathbf{M}}
\newcommand{\ubM}{\underline{\mathbf{M}}}

\newcommand{\ubI}{\underline{\mathbf{I}}}

\newcommand{\mR}{\mathcal{R}}

\newcommand{\rp}{r_{\mathrm{p}}}
\newcommand{\ra}{r_{\mathrm{a}}}

\newcommand{\bI}{\mathbf{I}}

\newcommand{\mJ}{\mathcal{J}}
\newcommand{\mI}{\mathcal{I}}

\newcommand{\ba}{\mathbf{a}}
\newcommand{\bb}{\mathbf b}

\newcommand{\bmom}{\boldsymbol{[\mu]}}
\newcommand{\uba}{\underline{\mathbf a}}
\newcommand{\ubb}{\underline{\mathbf b}}

\newcommand{\ubmom}{\underline{\bmom}}
\newcommand{\ubMmom}{\underline{\bM[\mu]}}

\newcommand{\Rb}{R_{\mathrm b}}

\newcommand{\nmax}{n_{\mathrm{max}}}

\newcommand{\gtorder}{\mathrel{\raise.3ex\hbox{$>$}\mkern-14mu
             \lower0.6ex\hbox{$\sim$}}}
\newcommand{\ltorder}{\mathrel{\raise.3ex\hbox{$<$}\mkern-14mu
             \lower0.6ex\hbox{$\sim$}}}
\newcommand{\vmax}{v_{\max}}
\newcommand{\rmax}{r_{\max}}
\newcommand{\rmin}{r_{\min}}
\newcommand{\rbas}{r_{\mathrm{bas}}}
\newcommand{\umax}{u_{\max}}
\newcommand{\umin}{u_{\min}}
\newcommand{\vmin}{v_{\min}}
\newcommand{\vt}{v_{\mathrm{t}}}

\newcommand{\Mtot}{M_{\mathrm{tot}}}

\newcommand{\Etr}{E_{\mathrm{tr}}}

\newcommand{\Ep}{E^{\prime}}
\newcommand{\Lp}{L^{\prime}}

\newcommand{\rext}{\mathrm{ext}}
\newcommand{\oFf}{\overline{F}_{\rf}}

\shorttitle{Collisionless relaxation near equilibrium}
\shortauthors{Rozier \& Errani}
\graphicspath{{./}{}}

\begin{document}

\title{Collisionless relaxation from near equilibrium configurations: Linear theory and application to tidal stripping}

\correspondingauthor{Simon Rozier}
\email{simon.rozier@ed.ac.uk}

\author{Simon Rozier}
\affiliation{School of Mathematics and Maxwell Institute for Mathematical Sciences, University of Edinburgh, Kings Buildings, Edinburgh, EH9 3FD, UK}
\author{Rapha\"el Errani}
\affiliation{McWilliams Center for Cosmology, Department of Physics, Carnegie Mellon University, Pittsburgh, PA 15213, USA}
\affiliation{Universit\'e de Strasbourg, CNRS, Observatoire astronomique de Strasbourg, UMR 7550, F-67000 Strasbourg, France}

\begin{abstract}
Placed slightly out of dynamical equilibrium, an isolated stellar system quickly returns towards a steady virialized state. We study this process of collisionless relaxation using the matrix method of linear response theory. We show that the full phase space distribution of the final virialized state can be recovered directly from the disequilibrium initial conditions, without the need to compute the time evolution of the system. This shortcut allows us to determine the final virialized configuration with minimal computational effort. Complementing this result, we develop tools to model the system's full time evolution in the linear approximation. In particular, we show that moments of the velocity distribution can be efficiently computed using a generalized moment matrix. We apply our linear methods to study the relaxation of energy-truncated Hernquist spheres, mimicking the tidal stripping of a cuspy dark matter subhalo. Comparison of our linear predictions against controlled, isolated $N$-body simulations shows agreement at per cent level for the parts of the system where a linear response to the perturbation is expected. We find that relaxation generates a tangential velocity anisotropy in the intermediate regions, despite the initial disequilibrium state having isotropic kinematics. Our results also strengthen the case for relaxation depleting the amplitude of the density cusp, without affecting its asymptotic slope. Finally, we compare the linear theory against an $N$-body simulation of tidal stripping on a radial orbit, confirming that the theory still accurately predicts density and velocity dispersion profiles for most of the system. 
\end{abstract}


\section{Introduction} 
\label{sec:intro}
On the long term, stellar systems evolve through a succession of dynamical (quasi-) equilibria. This evolution can either be driven by internal processes (such as secular relaxation or instabilities), or be externally induced (like, e.g., through tidal forces, or the accretion of mass). 
Analytical methods that model the return of a stellar system towards equilibrium in response to a perturbation can be broadly categorized into two classes, depending on whether the perturbation is assumed to act on a timescale that is long with respect to the system's internal dynamics, or short. 

For the case of slow perturbations, \cite{Young1980} developed a method based on the \textit{adiabatic} invariance of the orbital actions to study the response of a star cluster to the (slow) mass growth of a central black hole. \cite{Sellwood+2005} applied the same formalism to the contraction of a dark matter (DM) halo due to a growing baryonic disk, and more recently \cite{Stuecker+2023} used it to model the tidal evolution of DM subhalos. The \cite{Young1980} method allows for a computationally less costly calculation of the new equilibrium configuration than an $N$-body simulation would. However, the underlying assumptions do limit the range of problems that the method can be applied to: First, the assumption of a slow perturbation suppresses any reference to time evolution in the computation of the final state, so that there is no way to understand how the system reaches its final equilibrium. Second, the adiabatic invariance of the actions is not based on a solid theoretical background \citep{Weinberg1994AdiabI,Weinberg1994AdiabII}, and may be challenged as soon as one departs from the purely spherically symmetric system, i.e. as soon as the \textit{quasi-}periodic nature of the orbits comes into play\footnote{We note that in \cite{Stuecker+2023}, the method based on adiabatic conservation of the actions is much more accurate when the tide is spherically symmetric than when it is not.}. 

Methods developed to study the effect of a very fast perturbation with respect to the system's internal dynamics are generally based on the \textit{impulse} approximation. 
These methods consider that the perturbation transfers an instantaneous velocity kick to each particle in the system \citep[see e.g.][]{Kundic1995,GnedinHernquistOstriker1999}, which then settles into a new equilibrium, following some analytical prescriptions (such as, e.g., the conservation of the energy of individual shells of material in \citealt{Dutton2016}). Methods that make use of the impulse approximation have been used to 
estimate the effect of feedback-driven outflows on the central structure of DM halos \citep[e.g.][]{PontzenGovernato2012, Freundlich+2020, Li+2023}, as well to model the tidal evolution of galaxies and DM subhaloes \citep[e.g.][]{Gnedin+1999,TaylorBabul2001,Drakos+2020,Benson+2022}. 
Not all parameters for these models follow directly from first principles, but some require calibration to simulation results (e.g., the adiabatic parameter of \citealt{Gnedin+1999}, or the diffusion parameter of \citealt{Benson+2022}). 

The application of either the impulse or the adiabatic approximation becomes challenging as soon as the perturbed system spans a wide range of dynamical timescales, causing some parts of the system to perceive a given perturbation as adiabatic, and others as impulsive \citep[e.g.][]{Weinberg1994AdiabI,Weinberg1994AdiabII,Gnedin+1999}, 

In the present study, we develop an analytical method to model the process of collisionless relaxation for the specific case of weak perturbations to an initial equilibrium configuration. In this context, our method overcomes some of the limitations of existing tools, while still benefiting from the advantages of an analytical framework. Our model is rooted in the \citet{Kalnajs1977} matrix method of linear response theory (LRT), which allows us to explicitly solve for the evolution of the phase space distribution function (DF) at linear order in the perturbation, via the linearisation of the collisionless Boltzmann -- Poisson system of equations. 

As an example, we apply our method to study the response of \citet{Hernquist1990} spheres with initially isotropic kinematics to energy-based mass removal (similar to the initial conditions studied in \citealt{Drakos+2020}, \citealt{Amorisco2021}, \citealt{Errani+2022}, \citealt{Stuecker+2023}), approximating the type of perturbation typical for weak tidal encounters of cold DM subhaloes.
A theoretical understanding of the detailed (phase space) structure of tidally stripped subhaloes is of relevance, for example, for the the modelling of Milky Way dwarf galaxies \citep[see, e.g.,][]{Penarrubia+2008,Penarrubia+2010}, the search for dark substructures through strong gravitational lensing \citep[e.g.][]{Vegetti2009,Vegetti2010,Despali2022}, as well as the modelling of potential DM annihilation or decay signals \citep[e.g.][]{Goerdt2007, Stref2019, Facchinetti2022, Stuecker+2023Annihilation, Delos2023, Lovell2024}. 

The detailed evolution of the inner regions of cold DM haloes  under the effect of tides remains a matter of active debate. 
Multiple studies that use $N$-body simulations to model the tidal stripping of cuspy subhaloes strongly suggest a depletion in the amplitude of the central density cusp \citep[e.g.][]{Hayashi+2003, Penarrubia+2008, Penarrubia+2010, Green+2019, Errani+2021}, while recent studies based on analytical arguments suggest that the very center should remain virtually unaffected by tides \citep{Drakos+2020,Drakos2022,Stuecker+2023}. Motivated by these conflicting predictions, we will address the tidal evolution of the central density cusp using LRT. 

The paper is structured as follows. First, we show that we can cast the collisionless Boltzmann equation (CBE) so as to directly infer the characteristics of the final state from the initial conditions, in a single computational step. This allows us to compute the DF in the final equilibrium with minimal computational effort. The method is presented in Sec.~\ref{sec:Single_Computation}, which also includes a comparison against the results of controlled $N$-body simulations. Second, in Sec.~\ref{sec:Evolution}, we turn our attention to the time evolution of the system, and develop a method to efficiently compute the velocity moments of the evolving DF. Again, we compare the linear model predictions against controlled $N$-body simulations. Finally, in Sec.~\ref{sec:Discussion}, we discuss the application of this model to the relaxation of cuspy DM subhalos subject to tidal stripping. We show that the model is particularly suitable to study the tidal evolution of the very inner regions of a DM cusp, and that collisionless relaxation drives the depletion in amplitude of the DM cusp during tidal stripping. Finally, we summarize our main results and conclusions in Sec.~\ref{sec:Conclusion}.

\section{The new equilibrium in a single computation} 
\label{sec:Single_Computation}

\subsection{Model for a weakly perturbed equilibrium} 
\label{subsec:Model_Outline}

We aim at analysing a relatively broad class of physical problems, in which a self-gravitating system initially close to an equilibrium relaxes to reach a steady state. We specify here the assumptions that we make on the initial state out of equilibrium.

Our first assumption, that the system is initially close to an equilibrium, implies that there exists a decomposition of the initial phase space DF $F_{\ri}$ and the corresponding gravitational potential $\psi_{\ri}$ into a sum,
\begin{align}
\nonumber
    F_{\ri}(\bx, \bv) & = F_0(\bx, \bv) + u(\bx, \bv), \\
    \psi_{\ri}(\bx) & = \psi_0(\bx) + \psi_u(\bx),
    \label{eq:IC_Disequilibrium}
\end{align}
where $F_0$ is the DF of an equilibrium configuration\footnote{By convention, we normalize the DF to the total mass of the equilibrium,
\begin{equation}
    \!\! \int \!\! \rd \bx \rd \bv \, F_0(\bx, \bv) = \Mtot.
\end{equation}} with $\psi_0$ its gravitational potential, and $u$ represents a perturbation of the DF on top of this equilibrium, $|u| \ll F_0$, generating a potential $\psi_u$. We also introduce a small external perturbing potential $\psi_{\mathrm{ext}}$, with $|\psi_{\mathrm{ext}}| \ll |\psi_0|$, which is allowed to vary in time and is added on top of $\psi_{\ri}$. Note that the decomposition of $F_{\ri}$ is not unique, given that an equilibrium is usually surrounded by other equilibria in a continuous way. Here, we assume the potential $\psi_0$ to be integrable, hence we can define a set of angle-action variables $(\bT, \bJ)$ and apply Jeans' theorem \citep{BinneyTremaine2008}, so that $F_0$ is only a function of $\bJ$. We further assume that one of these decompositions lets us express $u$ too as a function of $\bJ$ only. This assumption means that in one of these equilibria, the perturbation that relates the equilibrium to our disequilibrium DF can be described as independent of the orbital phase. We can then define a new function $g = 1 + u / F_0$, so that
\begin{equation}
    F_{\ri}(\bJ) = g(\bJ) F_0(\bJ).
    \label{eq:Perturbation_Phase_Mixed}
\end{equation}
Assuming that the external perturber $\psi_{\mathrm{ext}}$ is asymptotically time stationary, the initial disequilibrium configuration $F_{\ri}$ is expected to reach a new steady state\footnote{Implicit here is the fact that the system does not support any neutral mode, which could theoretically drive it into infinite oscillations.}. In the following sections, we show that this final state can be efficiently computed using LRT.

These assumptions on the initial state still describe a relatively broad class of physical problems. We can consider the following scenario: a system in dynamical equilibrium undergoes a change of its potential, from an external source, which then remains stationary in time. In that case, $u=0$, $\psi_u = 0$ and the external perturber is represented by $\psi_{\rext}$. This scenario can describe a rapid infall of gas in a galaxy, or on the contrary a fast loss of mass due to stellar feedback in a young star cluster. Another scenario, which we will follow more closely in this paper, is that of a change in the equilibrium DF. In that case, $\psi_{\rext} = 0$ and the change in the DF is represented by $u$. This model can represent a system that is still going through the consequences of a past perturbation, once the perturber has disappeared. We will show in Sec.~\ref{subsec:Tidal_Stripping} that this model can describe the way a DM subhalo evolves after tidal stripping.

\subsection{Linear response theory} 
\label{subsec:Linear_Theory_Single_Step}

\subsubsection{A Boltzmann equation for the final state} 
\label{subsubsec:Boltzmann_Final_State}

In order to model the revirialization process from the initial state $(F_{\ri}, \psi_{\ri})$, we rely on the linearization of the Vlasov-Poisson system of equations. Here, we show that in our case of interest, computing the final equilibrium can be done very efficiently.

When the initial conditions of eq.~\eqref{eq:IC_Disequilibrium} are evolved in time, the initial DF and potential turn into a new state given by 
\begin{align}
    \nonumber
    F_1(\bx, \bv, t) & = F_{\ri}(\bx, \bv) + f(\bx, \bv, t), \\
    \psi_1(\bx, t) & = \psi_{\ri}(\bx) + \psi_{\mathrm{ext}}(\bx, t) + \psi^{\rs}(\bx, t),
    \label{eq:DF_Evolution_Definition}
\end{align}
where $f$ represents the evolution of the DF and $\psi^{\rs}$ the corresponding evolution of the potential. These responses in the DF and in the potential are related through 
\begin{equation}
    \psi^{\rs}(\bx, t) = - G \!\! \int \!\! \rd \bxp \, \rd \bvp \, \frac{f(\bxp, \bvp, t)}{|\bxp - \bx|}.
\end{equation}
The dynamical evolution of the system is then described by the Collisionless Boltzmann equation (CBE), which in an inertial reference frame can be written as
\begin{align}
    \nonumber
        \frac{\partial (F_{\ri} + f)}{\partial t} \,+\,& \frac{\partial (F_{\ri} + f)}{\partial \bT} \cdot \frac{\partial (H_0 + \psi^{\re} + \psi^{\rs})}{\partial \bJ} \\ 
        \,-\,& \frac{\partial (F_{\ri} + f)}{\partial \bJ} \cdot \frac{\partial (H_0 + \psi^{\re} + \psi^{\rs})}{\partial \bT} = 0,
\end{align}
where $H_0 = \tfrac{v^2}{2} + \psi_0(r)$ is the Hamiltonian in the equilibrium potential $\psi_0$, and $\psi^{\re} = \psi_u + \psi_{\rext}$. We can simplify this equation, as we know that neither $H_0$ nor $F_{\ri}$ depend on the angles $\bT$ (since $H_0$ is conserved along the orbits, and because of eq.~\eqref{eq:Perturbation_Phase_Mixed}), and that $F_{\ri}$ is static in time. We also apply here the linear approximation, considering that $|f| \ll F_{\ri}$ and $|\psi^{\re}|, |\psi^{\rs}| \ll |\psi_0|$ and neglecting second order terms. Noting $\bO(\bJ) = \partial H_0 / \partial \bJ$ the vector of orbital frequencies, the revirialization is consequently described by the linearized CBE, 
\begin{equation}
    \frac{\partial f}{\partial t} + \bO(\bJ) \cdot \frac{\partial f}{\partial \bT} - \frac{\partial F_{\ri}}{\partial \bJ} \cdot \frac{\partial (\psi^{\re} + \psi^{\rs})}{\partial \bT} = 0.
    \label{eq:Linearized_CBE}
\end{equation}
This equation, together with the linearized Poisson equation $\nabla^2 \psi^{\rs} = 4 \pi G \rho^{\rs}$, allow for the computation of the full revirialization process in the linear regime (see Sec.~\ref{sec:Evolution}).

In this section, we are merely interested in the final state after revirialization, and not in the details of the time evolution. Knowing that this final state is time stationary, it can be found by considering the time asymptotic limit of eq.~\eqref{eq:Linearized_CBE}, when $\partial f_{\rf} / \partial t = 0$, the subscript $\rf$ denoting the characteristics of the final equilibrium. Hence we have
\begin{equation}
    \bO(\bJ) \cdot \frac{\partial f_{\rf}}{\partial \bT} - \frac{\partial F_{\ri}}{\partial \bJ} \cdot \frac{\partial (\psi^{\re}_{\rf} + \psi^{\rs}_{\rf})}{\partial \bT} = 0.
    \label{eq:CBE_Asymptotic}
\end{equation}
It should be noted that $(\bT, \bJ)$ are still the angle-action coordinates in the equilibrium $\psi_0$. 

It is interesting to see that the final state can be computed, independently of the specific path that the system took to reach it. In particular, it does not depend on the way the perturbation $\psi_{\mathrm{ext}}$ is turned on, but only on its time asymptotic value. Consequently, in the linear approximation, the structure of the final equilibrium does not depend on whether the perturber is turned on impulsively or adiabatically. This may explain why collisionless relaxation in the linear limit can be studied within both the fast \citep{Amorisco2021} and the slow \citep{Stuecker+2023} approximations. Two remarks can be made about this finding. First, remaining in the linear limit implies that the perturbation $\psi_{\mathrm{ext}}$ is always small, including at intermediate times, and that the system's response does not involve non linear resonant interaction, in particular interaction with neutral or weakly damped modes \citep{Weinberg1994,Heggie+2020}. Second, the fact that the final state is independent from the actual time evolution is true only at the level of the global properties, i.e. the total DF and derived quantities. The fate of individual particles, however, may depend on how the perturber forces the system. \cite{Li+2023} illustrate that behaviour: they show that when the perturber is instantaneous, the action distribution after relaxation can be conserved even if the action of each individual particle is not.

\subsubsection{Final potential} 
\label{subsubsec:Final_Potential}

To solve the CBE--Poisson system of equations, we use the matrix method, as devised by \cite{Kalnajs1977}, which we adapt here to the specificities of eq.~\eqref{eq:CBE_Asymptotic}. First, for any integer vector $\bn$, we multiply eq.~\eqref{eq:CBE_Asymptotic} by $\re^{-\ri \bn \cdot \bT}$ and integrate over the angles, to get
\begin{equation}
    f_{\rf,\bn}(\bJ) = \frac{\bn \cdot \partial F_{\ri} / \partial \bJ}{\bn \cdot \bO} (\psi^{\re}_{\rf,\bn} + \psi^{\rs}_{\rf,\bn}),
\label{eq:CBE_Asymptotic_Angle_Averaged}
\end{equation}
where we defined the angle Fourier transform of any function $h$ of phase space to be
\begin{equation}
    h_{\bn}(\bJ) = \!\! \int \!\! \frac{\rd \bT}{(2 \pi)^3} \, h(\bT, \bJ) \, \re^{- \ri \bn \cdot \bT}.
    \label{eq:Angle_Fourier_Transform}
\end{equation}

Then, the perturbing potentials and densities are projected on a bi-orthogonal basis with potential elements $\psi^{(p)}$ and density elements $\rho^{(p)}$ related through the Poisson equation, so that
\begin{align}
\nonumber
    \psi^{\rs}(\bx) & = \sum_p a_p \psi^{(p)}(\bx), \\
    \psi^{\re}(\bx) & = \sum_p b_p \psi^{(p)}(\bx),
    \label{eq:BOB_Deprojection}
\end{align}
with the bi-orthogonality condition 
\begin{equation}
    \int \!\! \rd \bx \, \psi^{(p)*}(\bx) \, \rho^{(q)}(\bx) = - \delta_p^q.
\end{equation}
Following these definitions, the projections can be computed as, e.g., 
\begin{equation}
    b_{\rf,p} = - \!\! \int \!\! \rd \bx \, \psi^{\re}_{\rf}(\bx) \rho^{(p)*}(\bx) = - \!\! \int \!\! \rd \bx \, \rho^{\re}_{\rf}(\bx) \psi^{(p)*}(\bx).
    \label{eq:BOB_Projection}
\end{equation}
When applied to the final response with coefficients $a_{\rf,p}$, the projection gives
\begin{align}
\nonumber
    a_{\rf,p} &= - \!\! \int \!\! \rd \bx \, \rho_{\rf}^{\rs}(\bx) \, \psi^{(p)*}(\bx) \\
    &= - \!\! \int \!\! \rd \bx \rd \bv \, f_{\rf}(\bx, \bv) \, \psi^{(p)*}(\bx),
    \label{eq:Response_Asymptotic_Early_Demo}
\end{align}
where we used the fact that 
\begin{equation}
    \rho_{\rf}^{\rs}(\bx) = \!\! \int \!\! \rd \bv \, f_{\rf}(\bx, \bv).
\end{equation}
From eq.~\eqref{eq:Angle_Fourier_Transform}, and because any function of phase space is $2\pi$-periodic in the angles, the Fourier series of $f_{\rf}$ follows the convention
\begin{equation}
    f_{\rf}(\bT, \bJ) = \sum_\bn f_{\rf, \bn}(\bJ) \, \re^{\ri \bn \cdot \bT},
    \label{eq:Fourier_Transform}
\end{equation}
which we can inject into eq.~\eqref{eq:Response_Asymptotic_Early_Demo}. We can now canonically exchange integration variables from $\rd \bx \rd \bv$ to $\rd \bT \rd \bJ$, with a Jacobian equal to 1 due to phase space volume conservation. Therefore, we have
\begin{align}
\nonumber
    a_{\rf,p} &= - \sum_{\bn} \!\! \int \!\! \rd \bJ f_{\rf, \bn}(\bJ) \bigg( \!\! \int \!\! \rd \bT \, \re^{-\ri \bn \cdot \bT} \psi^{(p)}(\bT, \bJ) \bigg)^* \\
    & = - (2\pi)^3 \sum_{\bn} \!\! \int \!\! \rd \bJ \, f_{\rf, \bn}(\bJ) \psi_{\bn}^{(p)*}(\bJ),
\label{eq:Response_Asymptotic_Intermediate_Demo}
\end{align}
where $\psi^{(p)}(\bT, \bJ) = \psi^{(p)}(\bx(\bT, \bJ))$. We can now use eq.~\eqref{eq:CBE_Asymptotic_Angle_Averaged} to get
\begin{equation}
    a_{\rf,p} \!=\! -(2 \pi)^3 \!\sum_{\bn}\! \!\! \int \!\!\! \rd \bJ \frac{\bn \!\cdot\! \partial F_{\ri} / \partial \bJ}{\bn \cdot \bO} \psi_{\bn}^{(p)*}\!(\bJ) (\psi^{\re}_{\rf,\bn}\!(\bJ) \!+\! \psi^{\rs}_{\rf,\bn}\!(\bJ)).
\end{equation}

Finally, we can expand the perturbing potentials in the last parenthesis onto the bi-orthogonal basis following eq.~\eqref{eq:BOB_Deprojection}, yielding
\begin{align}
\nonumber
    a_{\rf,p} = -(2 \pi)^3 \sum_q \!\sum_{\bn} \!\! \int \!\! \rd \bJ & \frac{\bn \!\cdot\! \partial F_{\ri} / \partial \bJ}{\bn \cdot \bO} \psi_{\bn}^{(p)*}\!(\bJ)\psi_{\bn}^{(q)}(\bJ) \\ & \times (b_{\rf,q} + a_{\rf,q}).
\end{align}
This equation is a matrix equation,
\begin{equation}
    \ba_{\rf} = \bM_{\rf} (\ba_{\rf} + \bb_{\rf}),
    \label{eq:Matrix_Simple_Relation}
\end{equation}
where the matrix $\bM_{\rf}$ is defined by
\begin{equation}
    \bM_{\rf, pq} \! = \! - (2 \pi)^3 \! \sum_{\bn} \!\!\int\!\! \rd \bJ \, \frac{\bn \!\cdot\! \partial F_{\ri}/\partial \bJ}{\bn \cdot \bO} \psi_{\bn}^{(p)*\!}(\bJ) \psi_{\bn}^{(q)\!}(\bJ).
    \label{eq:Response_Matrix_Asymptotic}
\end{equation}
We can also rearrange eq.~\eqref{eq:Matrix_Simple_Relation} to show the linear relation between the perturber and the response,
\begin{equation}
    \ba_{\rf} = ([\bI - \bM_{\rf}]^{-1} - \bI) \bb_{\rf}.
    \label{eq:Linear_Equation_Asymptotic}
\end{equation}

This relation between the time asymptotic response and the perturber highlights again the fact that the final equilibrium that the system reaches is already encoded in the initial conditions, and is therefore independent from the details of the system's temporal evolution, in the linear approximation. In other words, a transient linear perturbation cannot have an ever-lasting effect on the system.

\subsubsection{Final phase space DF} 
\label{subsubsec:Final_DF}

From a computation of the final potential, we can recover the full phase space DF of the final equilibrium, $F_{\rf} = F_{\ri} + f_{\rf}$. A priori, it can be obtained by combining eqs.~\eqref{eq:Fourier_Transform} and~\eqref{eq:CBE_Asymptotic_Angle_Averaged} to get
\begin{equation}
    F_{\rf}(\bTp,\bJp) \!=\! F_{\ri}(\bJ) \!+\!\! \sum_\bn \frac{\bn \cdot \partial F_{\ri} / \partial \bJ}{\bn \cdot \bO(\bJ)} \big[\psi^{\re}_{\rf,\bn}(\bJ) + \psi^{\rs}_{\rf,\bn}(\bJ)\big] \, \re^{\ri \bn \cdot \bT}\!\!,
\end{equation}
where $(\bT,\bJ)$ are the angle-action coordinates computed in the potential $\psi_0$, of the same phase space point of angle-action coordinates $(\bTp,\bJp)$ when computed in the final potential $\psi_{\rf} = \psi_0+\psi^{\re}_{\rf}+\psi^{\rs}_{\rf}$. We additionally know that the final state is a new equilibrium, so that it should be independent of the phase angles $\bTp$. Because of the linear approximation, full phase mixing of the final state is only approximately reached by LRT. In order to improve the quality of our reconstruction of the final equilibrium, we force phase mixing in the final state by considering the final DF to be
\begin{equation}
    \oFf(\bJp) = \!\! \int \!\! \frac{\rd \bTp}{(2\pi)^3} F_{\rf}(\bTp, \bJp).
    \label{eq:Phase_Mixed_DF}
\end{equation}
This DF can be used, together with the final potential $\psi_{\rf}$, to compute velocity moments in the final state.

\subsection{Comparison of the final state with $N$-body simulations} 
\label{subsec:Simulations_Single_Step}

In order to validate our theory for collisionless relaxation close to an equilibrium, we compare the outcome of linear theory to $N$-body simulations, in a set of near-equilibrium systems which we define in the following section.

\subsubsection{Model for a near-equilibrium system} 
\label{subsubsec:Model_Near_Equilibrium}

\begin{figure}
    \centering
    \includegraphics[width=\columnwidth]{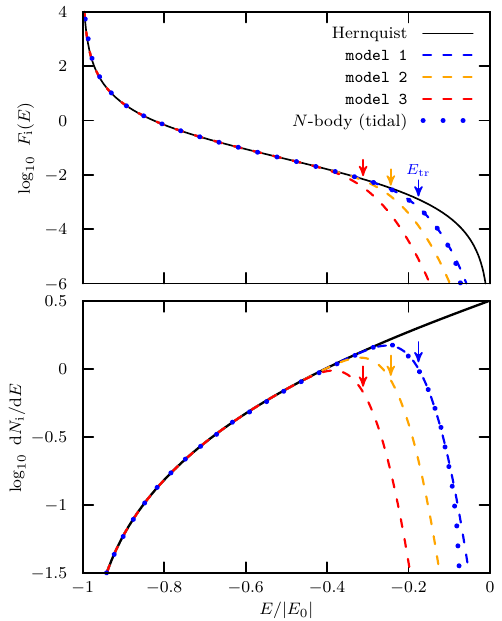}
    \caption{Top panel: Phase space \DF\@ of the ergodic Hernquist sphere $F_0$ (black, Eq.~\ref{eq:Hernquist_DF}), and of three near-equilibrium models $F_{\ri}$ (\texttt{model\,1}, \texttt{2}, \texttt{3} in blue, orange and red respectively, see Eq.~\ref{eq:Perturbation_Phase_Mixed} and Eq.~\ref{eq:truncation_function} using parameters as listed in Table~\ref{tab:parameter_overview}), as a function of the orbital energy $E$ (defined in the original Hernquist potential). The DFs are expressed in units of $\Mtot r_{\rs}^{-3}(-E_0)^{-3/2}$. 
    Arrows indicate the respective energy truncation scale $E_\mathrm{tr}$ (Eq.~\ref{eq:truncation_function}) for the three different models.
    Bottom panel: Initial differential energy distributions $\mathrm{d}N_\mathrm{i}/\mathrm{d}E$ (Eq.~\ref{eq:dNdE}) corresponding to the cases shown above. In each panel, blue dotted curves show the energy-truncated (disequilibrium) \DF\@ and energy distribution measured from an $N$-body simulation of a Hernquist subhalo in a tidal field (see Sec.~\ref{sec:MatrixTidal} for details).}
\label{fig:DF_Hernquist}
\end{figure}

We construct our initial DF $F_{\ri}$ in  the following way: first, we choose the equilibrium DF-potential pair, $F_0-\psi_0$; then we choose the function $g$ of eq.~\eqref{eq:Perturbation_Phase_Mixed}, so that the hypotheses of the model are verified. We consider an equilibrium following the Hernquist potential \citep{Hernquist1990}, i.e. with 
\begin{equation}
    \psi_0(r) = -\frac{G \Mtot}{r + r_{\rs}},
\end{equation}
where $\Mtot$ is the total mass of the model and $r_{\rs}$ its scale radius. Furthermore, we consider the equilibrium DF to be ergodic, i.e.
\begin{align}
\nonumber
    F_0(E) = & \frac{\Mtot}{8 \sqrt{2} \pi^3 r_{\rs}^3 (-E_0)^{3/2}} \Bigg[ \frac{3 \arcsin \!\Big(\! \sqrt{\tfrac{E}{E_0}} \!\Big)\!}{ \big( 1-\tfrac{E}{E_0}\big)^{5/2}} \\
    & + \frac{\sqrt{\tfrac{E}{E_0}} \big(1-2\tfrac{E}{E_0}\big) \big(8(\tfrac{E}{E_0})^2 - 8\tfrac{E}{E_0} - 3 \big)}{(1-\tfrac{E}{E_0})^2} \Bigg],
    \label{eq:Hernquist_DF}
\end{align}
where the scale energy is given by 
\begin{equation}
    E_0 = \psi_0(0) = - \frac{G \Mtot}{r_{\rs}}.
\end{equation}
This DF is represented as a black line in the top panel of Fig.~\ref{fig:DF_Hernquist}. The corresponding differential energy distribution $\mathrm{d}N / \mathrm{d}E$ can be obtained by multiplying eq.~\eqref{eq:Hernquist_DF} by the density of states $G_0(E)$,
\begin{equation}
    \mathrm{d}N / \mathrm{d}E = F_0(E) ~ G_0(E)
    \label{eq:dNdE}
\end{equation}
where
\begin{align}
\nonumber
&G_0(E) = \frac{2 \sqrt{2} \pi^2 r_{\rs}^{5/2} \sqrt{GM_\mathrm{tot}} }{3 (\tfrac{E}{E_0})^{5/2}} ~ \Big[ 3 \left(  8 (\tfrac{E}{E_0})^2 - 4\tfrac{E}{E_0} +1 \right)  \\
&\times \arccos \!\Big(\!\sqrt{\tfrac{E}{E_0}}\!\Big)\! - \sqrt{\tfrac{E}{E_0} - (\tfrac{E}{E_0})^2} (4\tfrac{E}{E_0} -1 ) (2 \tfrac{E}{E_0} +3) \Big].
\end{align}
This differential energy distribution is represented as a black line in the bottom panel of Fig.~\ref{fig:DF_Hernquist}. 

To perturb this equilibrium, we consider a function $g$ of the following form\footnote{The function $g(E)$ takes the role of the ``filter function'' in \citet{Errani+2022}, see their fig.~4 and eq.~9. We opt for a different parametrization which is better suited to approximate the shape of the tidal energy truncation typical for mildly perturbed Hernquist spheres, see Sec.~\ref{sec:MatrixTidal}.},
\begin{equation}
\label{eq:truncation_function}
    g(E) = \frac{1}{2} \, \mathrm{erfc} \bigg(\! \frac{E - \Etr}{A}  \!\bigg), 
\end{equation}
where erfc is the complementary error function. When $E = E_0$, we typically have $g \simeq 1$, and typically $g \simeq 0$ when $E=0$. This family has two free parameters, $\Etr$ and $A$, which respectively control the position in energies where the transition from 1 to 0 occurs, and the width of this transition region. This behaviour is illustrated in Fig.~\ref{fig:DF_Hernquist}, where we show the DF and differential energy distribution of three cases with $E_\mathrm{tr}/|E_0|=-0.176$, $-0.244$, and $-0.312$ (\texttt{model\,1}, \texttt{2} and \texttt{3}, see Tab.~\ref{tab:parameter_overview}). For all models considered in this study, we use the same value of $A/|E_0|=0.077$. 

Physically, the equilibrium is perturbed in the following way: nearly all orbits with energies below the transition region are conserved in the perturbed initial state, while hardly any orbit with energy above the transition region is kept. This is aimed at modelling a system that has lost its high energy orbits instantaneously due to an external process, which we show relates to the tidal stripping of DM substructure in Sec.~\ref{subsec:Tidal_Stripping}. We already see from Fig.~\ref{fig:DF_Hernquist} that the linear approximation is not verified in parts of the phase space with high energy, a point which we discuss in Sec.~\ref{subsubsec:Isolated_NBody_FinalState_results}.

\begin{table}
    \centering
    \caption{Initial conditions of the energy truncated Hernquist models considered in this study. We denote by $E_\mathrm{tr}$ the truncation energy as defined in Eq.~\ref{eq:truncation_function} (with $A/|E_0|=0.077$ for all three model), and by $M_\mathrm{tr}$ and $M_\mathrm{tot}$ the total masses of the truncated model and the original Hernquist model, respectively. The third column lists the mass enclosed within the Hernquist scale radius $r_\mathrm{s}$ normalized by the enclosed mass in the original Hernquist model $M_\mathrm{H}(<r_\mathrm{s})=0.25\,M_\mathrm{tot}$, }
    \begin{tabular}{l@{\hskip 0.8cm}c@{\hskip 0.8cm}c@{\hskip 0.4cm}c}
    \toprule
                                       & $E_\mathrm{tr}/|E_0|$ & $M_\mathrm{tr}/M_\mathrm{tot}$ & $M_\mathrm{tr}(<r_\mathrm{s})/(0.25\,M_\mathrm{tot})$   \\ \midrule
       {\color{blue}\texttt{model~1}}  &   -0.176        & 0.56                           & 0.98    \\  
       {\color{orange}\texttt{model~2}}&   -0.244        & 0.43                           & 0.95    \\  
       {\color{red}\texttt{model~3}}   &   -0.312        & 0.33                           & 0.90    \\  
       \bottomrule
    \end{tabular}
    \label{tab:parameter_overview}
\end{table}

\subsubsection{Application of the matrix method} 
\label{subsubsec:Application_Matrix_FinalState}

Once the initial DF $F_{\ri}$ is defined, we apply the matrix method in the following steps:
\begin{itemize}
    \item Construct the perturbing density
    as
    \begin{equation}
        \rho^{\re}(\bx) = \rho_u(\bx) = \!\! \int \!\! \rd \bv \, (g(\bx,\bv) - 1) F_0(\bx,\bv).
        \label{eq:Perturbing_Density_Asymptotic}
    \end{equation}
    \item Project it onto the basis following eq.~\eqref{eq:BOB_Projection}, to get the vector $\bb_{\rf}$. For these projections, we used a separable spherical basis with spherical harmonics as the angular functions\footnote{Because of spherical symmetry, there is no actual angular dependence, i.e. the multipolar expansion is restricted to the first term, $\ell = m = 0$.}, and radial functions from \cite{CluttonBrock1973}, with $\nmax = 201$ (i.e., 202 basis functions) and characteristic radius $\Rb = 3 \, r_{\rs}$.
    \item Independently compute the response matrix of eq.~\eqref{eq:Response_Matrix_Asymptotic} from $F_{\ri}$. The action space integral considers all orbits with $0.002 \, r_{\rs} < \rp < \ra < 20 \, r_{\rs}$. The sum over angular Fourier numbers is performed with $n_2 = n_3 = 0$ (spherical symmetry of the background and the perturber) and $|n_1| \leq 10$. We confirmed the convergence of our results w.r.t. changes in these parameters.
    \item Apply eq.~\eqref{eq:Linear_Equation_Asymptotic} to get the vector $\ba_{\rf}$. Then, the response density and potential can be recovered by de-projection using eq.~\eqref{eq:BOB_Deprojection}.
    \item Recover the full phase space distribution using Eq.~\ref{eq:Phase_Mixed_DF}. Owing to spherical symmetry, the integral over $\bTp$ in that equation reduces to an integral over $\theta_1^{\prime}$. After a change of variables $\theta_1^{\prime} \to r$, we compute the $(v_r, \vt)$ coordinates of the phase space point of coordinates $(r, \Ep, \Lp)$. Then, the values of $(E, L)$ can be easily found, while that of $\theta_1$ is performed via a 1D integral \citep[see, e.g.,][]{TremaineWeinberg1984}. To compute moments of the velocity distribution, we interpolate the DF $\overline{F}_{\rf}$ on a grid of $(\Ep, \Lp)$, then we perform standard integrals of the interpolated function.
\end{itemize}
The bottleneck of this procedure is the computation of the response matrix. We use numerical methods straightforwardly adapted from \cite{Rozier2019}\footnote{Note the similarity between eq.~\eqref{eq:Response_Matrix_Asymptotic} and their eq.~(9).} and summarized in App.~\ref{subsec:Numerical_Details}, which we complemented with a method to include edge terms in the action space integration, as highlighted in App.~\ref{app:Edges}. Appendix~\ref{subsec:Complexity} discusses the numerical complexity of the calculation.  We show the results of the whole procedure in Sec.~\ref{subsubsec:Isolated_NBody_FinalState_results}.

\subsubsection{Isolated simulations setup} 
\label{subsubsec:Isolated_NBody_FinalState_setup}
We now briefly outline the numerical setup of the isolated $N$-body runs which we use to compare the matrix results against. 
Our simulation setup is strongly motivated by the type of controlled simulations studied in \citet{Amorisco2021}.

\textit{$N$-body models.} We generate an equilibrium $N$-body realization with $N=10^7$ particles of a spherical Hernquist model with isotropic velocity dispersion, using the implementation\footnote{Available at \url{https://github.com/rerrani/nbopy}.} of \cite{Errani+2020}. This initial sampling is represented as black dots in Fig.~\ref{fig:DF_Hernquist}.
In a subsequent step, we build three different disequilibrium $N$-body models by truncating the Hernquist realization in energy, following a tapered truncation as parameterized by Eq.~\ref{eq:truncation_function} with $A/|E_0|=0.077$ and $E_\mathrm{tr}/|E_0|=-0.176$, $-0.244$, and $-0.312$ (\texttt{model\,1}, \texttt{2} and \texttt{3}, see Tab.~\ref{tab:parameter_overview}).
These disequilibrium models are obtained by rejection sampling. They consist of $5.6\times10^6$ and $4.3\times10^6$ and $3.3\times10^6$ $N$-body particles, respectively. 

\textit{Particle-mesh code.} We compute the dynamical evolution of our $N$-body models using the particle-mesh code \textsc{superbox} \citep{Fellhauer+2000}. The code employs a high- and a medium resolution cubic grid with $256^3$ cells each, centred on and co-moving with the $N$-body model, with resolutions of $\Delta x_1 \approx 0.008\, r_\mathrm{s}$ and $\Delta x_2 \approx 0.08\, r_\mathrm{s}$, respectively. A third, static grid with a lower resolution of $\Delta x_3 \approx 0.8\, r_\mathrm{s}$ contains the entire simulation volume. The time integration makes use of a leapfrog scheme with fixed time step $\Delta t = T_\mathrm{s}/1600$, where 
\begin{equation}
T_\mathrm{s} = 4 \pi r_\mathrm{s}^{3/2} (G \Mtot)^{-1/2}
\label{Eq.:Ts}
\end{equation}
denotes the period of a circular orbit with radius $r_\mathrm{s}$ in the original Hernquist potential. We follow the virialisation of the disequilibrium $N$-body model for a total time of $15\,T_\mathrm{s}$. 

\begin{figure}
    \centering
    \includegraphics[width=\columnwidth]{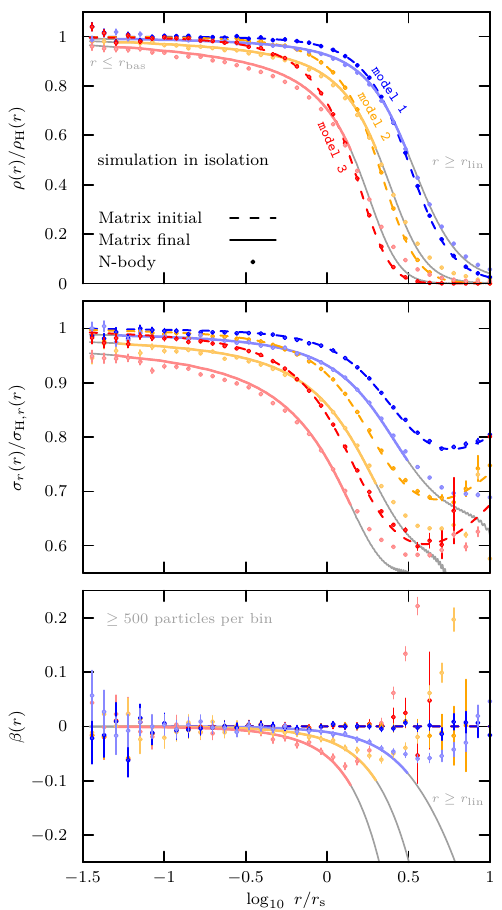}
    \caption{Top panel: Density $\rho(r)$ in the initial disequilibrium and final relaxed states, normalized by the Hernquist density $\rho_\mathrm{H}(r)$. Blue, orange and red dashed curves correspond to the energy-truncated initial conditions of \texttt{model\,1}, \texttt{2} and \texttt{3} with parameters as listed in Table~\ref{tab:parameter_overview}. Results of the matrix method are shown as solid curves, whereas the corresponding $N$-body simulations (at $t=15\,T_\mathrm{s}$) are shown as filled circles. Errorbars indicate the expected $N$-body Poisson noise in each bin. Middle panel: as before, but showing the radial velocity dispersion $\sigma_r(r)$. Bottom panel: as before, but showing the anisotropy parameter $\beta = 1 - \sigma^2_{\rt}/(2 \sigma^2_r)$.}
    \label{fig:final_state_comparison}
\end{figure}

\subsubsection{Comparison of the final state between matrix and isolated simulations} 
\label{subsubsec:Isolated_NBody_FinalState_results}
Figure~\ref{fig:final_state_comparison} compares the final relaxed states computed with the matrix method against our $N$-body models, after evolving the latter in isolation for a duration of $t=15\,T_\mathrm{s}$. The $N$-body models are shown as filled circles, with errorbars indicating the level of Poisson noise. Analytical initial values, as well as the results of calculations performed using the matrix method are shown as solid curves.

Dashed curves in the top panel of Fig.~\ref{fig:final_state_comparison} show the initial density profiles of the three disequilibrium models in blue, orange and red (for \texttt{model\,1}, \texttt{2} and \texttt{3}, respectively, see Tab~\ref{tab:parameter_overview}). Densities are normalized by the Hernquist density, $\rho_{\mathrm{H}}$. As expected, the truncation in energy introduces a depletion in density in the outer regions. In the inner regions ($r \lesssim r_\mathrm{s}$), the initial disequilibrium profiles approach the original Hernquist density. 
The corresponding final equilibrium density profiles are shown as solid curves of lighter shade. In the final equilibrium state, the inner density is slightly depleted compared to the initial disequilibrium state (the behaviour in the very centre is discussed in Sec.~\ref{subsec:Central_Density}), whereas the density in the outer regions slightly increases. At radii where the results of the matrix method are likely affected by resolution limits of the underlying basis ($r \leq r_\mathrm{bas}$), or where non-linear evolution is expected and the matrix method is not applicable ($r \geq r_\mathrm{lin}$), the matrix results are shown in grey. We choose $r_\mathrm{bas} = 3 \, \Rb / \nmax$, which is approximately the position of the second node of the last basis function (i.e., that with the highest radial resolution), and $r_\mathrm{lin}$ is the position at which $\rho^{\re}(r) = 0.5 \, \rho_{\mathrm{H}}$ in each case, i.e. already far from a linear perturbation. Remarkably, the density in the final state as computed with the matrix method is in excellent agreement with the results of our $N$-body simulations at all radii, including regions where the perturbation is strongly non-linear. 

We now turn our attention to the kinematics of the systems in question. The central panel of Fig.~\ref{fig:final_state_comparison} shows the radial velocity dispersion $\sigma_r$ of our models, in units of the isotropic Hernquist radial velocity dispersion, $\sigma_{\mathrm{H},r}$. As before, the initial disequilibrium systems are shown as dashed curves in red, orange and blue, while the final states are shown as solid curves in lighter shades of the same colours. The results of the matrix computation are in excellent agreement with our $N$-body models at those radii where the system is expected to respond linearly to the initial perturbation ($r \leq r_\mathrm{lin}$). At larger radii, the dispersion measured in the $N$-body models are slightly larger than suggested by the (linear) matrix calculation. From these curves, we clearly see that our case-dependent criterion for linearity of the perturbation is valid: the velocity dispersion computed from LRT starts to depart from the non-linear $N$-body profile as soon as, in each case, the perturber's density represents a dominant fraction of the initial density. 

Finally, we compare the anisotropy measured in the $N$-body models against the matrix calculations. We quantify the anisotropy through the usual anisotropy parameter $\beta = 1 - \sigma^2_{\rt}/(2 \sigma^2_r)$, where subscripts $\rt$ and $r$ denote tangential and radial components, respectively. Both the matrix calculation, and the $N$-body model show that the relaxed equilibrium system remains isotropic in the centre, but develops a slight tangential anisotropy ($0 \geq \beta \gtrsim -0.1$) at radii of $1 \lesssim r_\mathrm{s} \lesssim 3$. At larger radii, the matrix method is not applicable; the $N$-body models suggest that $\beta$ rises again at radii beyond a few $r_\mathrm{s}$.

To summarize, the results of the matrix calculation are in excellent agreement with those of controlled $N$-body simulations at radii where a linear response to the perturbation is expected ($r \leq r_\mathrm{lin}$). Remarkably, the density profile as computed with the matrix methods remain consistent with the $N$-body models also at radii beyond $r_\mathrm{lin}$.

\section{Evolution towards the relaxed state} 
\label{sec:Evolution}

As derived in the previous section, the final equilibrium configuration may be computed directly from the initial disequilibrium distribution, without explicitly modelling the system's time evolution. 
Complementing this result, we now turn our attention to model in detail how the equilibrium configuration is reached, and use the matrix method of LRT to predict the full time evolution of the system in response to a perturbation.

\subsection{Evolution of the density} 
\label{subsec:Density_Evolution}

The evolution of a perturbed equilibrium's density with LRT has already been developed in the literature \citep{Seguin1994,Weinberg1998,Murali1999,Pichon2006,Dootson+2022}, we briefly summarize the main procedure here, following \cite{Rozier+2022}. We start with the linearized CBE--Poisson system of partial differential equations (see eq.~\eqref{eq:Linearized_CBE} and the following paragraph). On the one hand, we multiply the CBE by $\re^{-\ri \bn \cdot \bT}$ and integrate over the angles, which gives us a differential equation for each set of Fourier numbers $\bn$. Each equation can be solved for $f_{\bn}(\bJ,t)$, yielding
\begin{equation}
    f_{\bn}(\bJ,t) \!=\! \ri \, \bn \!\cdot\! \frac{\partial F_{\ri}}{\partial \bJ} \!\! \int_0^t \!\!\!\! \rd \tau \, [\psi_{\bn}^{\re}(\bJ,\tau) + \psi_{\bn}^{\rs}(\bJ,\tau)] \re^{-\ri \bn \cdot \bO (t - \tau)}\!\!.
    \label{eq:Time_Evolution_DF_Fourier_Integral}
\end{equation}

On the other hand, we can project the response density $\rho^{\rs}(\bx,t)$ onto the same basis as defined in eq.~\eqref{eq:BOB_Deprojection}, with projection coefficients $a_p(t)$. The exact same mathematical steps as eqs.~\eqref{eq:Response_Asymptotic_Early_Demo} to~\eqref{eq:Response_Asymptotic_Intermediate_Demo} can be applied to $a_p(t)$, replacing the final state with the evolving state. Here, the Fourier transformed DF should be replaced with eq.~\eqref{eq:Time_Evolution_DF_Fourier_Integral}, yielding a matrix equation with a remaining time integral,
\begin{equation}
    \ba(t) = \!\! \int_0^t \!\!\! \rd \tau \, \bM(t - \tau) [\bb(\tau) + \ba(\tau)],
    \label{eq:Matrix_Time_Integral}
\end{equation}
where the response matrix is defined by
\begin{equation}
    \bM_{pq}(t) \!\! = \!\! - \ri (2 \pi)^3 \! \sum_{\bn} \!\!\int\!\! \rd \bJ \, \re^{- \ri \, \bn \cdot \bO \, t} \, \bn \!\cdot\! \frac{\partial F_{\ri}}{\partial \bJ} \psi_{\bn}^{(p)*\!}(\bJ) \psi_{\bn}^{(q)\!}(\bJ).
\label{eq:Response_Matrix}
\end{equation}

Finally, we can fully linearize the problem by discretizing the time integration. To that end, we fix a time interval $[0, T]$ for computing the evolution, we evenly divide it into $K+1$ steps $0=t_0 < ... < t_K = T$, and define new vectors $\uba$ and $\ubb$ obtained by stacking all vectors $\ba(t_0), ..., \ba(t_K)$ and $\bb(t_0), ..., \bb(t_K)$ respectively. With these definitions, \cite{Rozier+2022} show that the response $\uba$ is linearly related to the perturber $\ubb$ through 
\begin{equation}
    \uba = \big([\ubI - \ubM]^{-1} - \ubI\big) \, \ubb,
    \label{eq:Linear_response_final}
\end{equation}
where $\ubI$ is the identity matrix of suitable size, and the matrix $\ubM$ is defined by blocks, so that the block in the line $i$ and column $j$ (with $1 \leq i \leq K+1$ and $1 \leq j \leq K+1$) is given by
\begin{equation}
\ubM_{ij} = \left\{\begin{aligned}
&\Delta t \, \bM(t_i - t_j) & \, \mathrm{for} \, j < i, \\
& \mathbf{0} & \, \mathrm{for} \, j \geq i,
\end{aligned}\right.
\label{eq:Full_time_matrix}
\end{equation}
with $\Delta t = T/K$.

\subsection{Evolution of the velocity distribution} 
\label{subsec:Velocity_Evolution}

We are also interested in the evolution of some moments of the velocity distribution during revirialization, such as the mean radial velocity or the velocity dispersions. We note that we can recover the full phase space DF of the evolved system, using a method from \cite{Dury2008}. Here, we show that moments of the velocity distribution can be computed efficiently in the framework of the matrix method. 

Let us consider the computation of a moment $\langle \mu(\bv) \rangle(\bx, t)$ of the velocity distribution at a given position $\bx$ and a given time $t$ of the system's evolution. By definition, we have
\begin{equation}
    \langle \mu(\bv) \rangle(\bx, t) = \frac{1}{\rho(\bx, t)} \!\! \int \!\! \rd \bv \, \mu(\bv) \, F_1(\bx, \bv, t),
\end{equation}
where $\rho = \rho_0+\rho_u+\rho^{\rs}$ is the system's total density, excluding the \textit{external} part of the perturbation, and $F_1$ is defined by eq.~\eqref{eq:DF_Evolution_Definition}. The integral in the r.h.s can be decomposed into the initial contribution in $F_{\ri}$ and the evolutionary contribution in $f$. While the initial contribution can be straightforwardly computed from the initial DF, the evolutionary part requires knowledge of the full DF perturbation $f$, which is not as direct for now. So let us consider the term defined by
\begin{equation}
    \rho(\bx,t) \langle \mu \rangle_{\!f}(\bx,t) = \!\! \int \!\! \rd \bv \, \mu(\bv) \, f(\bx, \bv, t).
    \label{eq:Velocity_Moment_Evolutionary}
\end{equation}
Since it is a function of $\bx$ and $t$, it can be projected onto the bi-orthogonal basis of densities of eq.~\eqref{eq:BOB_Deprojection}, with projection coefficients $[\mu]_p(t)$ given by
\begin{equation}
    [\mu]_p(t) = - \!\! \int \!\! \rd \bx \, \rho(\bx,t) \langle \mu \rangle_{\!f}(\bx,t) \, \psi^{(p)*}(\bx).
\end{equation}
Replacing with eq.~\eqref{eq:Velocity_Moment_Evolutionary}, we have
\begin{equation}
    [\mu]_p(t) = - \!\! \int \!\! \rd \bx \, \rd \bv \, \mu(\bv) \, f(\bx, \bv, t) \, \psi^{(p)*}(\bx).
\end{equation}

Again, we can follow the same steps as eqs.~\eqref{eq:Response_Asymptotic_Early_Demo} to~\eqref{eq:Response_Asymptotic_Intermediate_Demo}, yielding
\begin{equation}
    [\mu]_p(t) = - (2 \pi)^3 \sum_{\bn} \!\! \int \!\! \rd \bJ \, f_{\bn}(\bJ, t) \, [\mu \psi^{(p)}]_{\bn}^*(\bJ).
    \label{eq:Moment_Coeff_Intermediate}
\end{equation}
When we inject eq.~\eqref{eq:Time_Evolution_DF_Fourier_Integral} into eq.~\eqref{eq:Moment_Coeff_Intermediate}, we see that it can be transformed into a similar matrix equation as eq.~\eqref{eq:Matrix_Time_Integral}, i.e.
\begin{equation}
    \bmom(t) = \!\! \int_0^t \!\! \rd \tau \, \bM[\mu](t - \tau) \big[ \bb(\tau) + \ba(\tau) \big],
    \label{eq:Moment_Matrix_Equation}
\end{equation}
where $\ba$ and $\bb$ are the same response and perturber vectors as in eq.~\eqref{eq:Matrix_Time_Integral}, $\bmom$ is the vector made of $[\mu]_p$, and $\bM[\mu]$ is the moment response matrix, defined by
\begin{align}
\nonumber
    \bM[\mu]_{pq}(t) = - \ri \, (2\pi)^3 \sum_{\bn} \!\! &\int \!\! \rd \bJ \, \re^{-\ri \bn \cdot \bO t} \, \bn \cdot \frac{\partial F_{\ri}}{\partial \bJ} \\
    & \times [\mu \psi^{(p)}]^*_{\bn}(\bJ) \, \psi_{\bn}^{(q)}(\bJ).
    \label{eq:Moment_Matrix}
\end{align}
Equation~\eqref{eq:Moment_Matrix_Equation} shows that the moment $\mu$ derives from integrating the dynamics in the fully perturbed potential $\ba + \bb$, keeping track of the value of the moment along the evolution through the moment response matrix $\bM[\mu]$. 

Finally, we can apply the same time discretization as in Sec.~\ref{subsec:Density_Evolution}, define the stacked vector $\ubmom$ in the same way as $\uba$ and $\ubb$, and the matrix $\ubMmom$ in the same way as eq.~\eqref{eq:Full_time_matrix}, to get
\begin{equation}
    \ubmom = \ubMmom (\uba + \ubb).
\end{equation}
This equation shows that the full time evolution of the function $\rho \, \langle\mu\rangle_{\!f}$, i.e. the density-weighted perturbation to the velocity moment, is linearly related to the total potential perturbation to the system, $\psi^{\re}+\psi^{\rs}$, through the moment response matrix $\ubMmom$.

\subsection{Application of the matrix method} 
\label{subsec:Application_Matrix_Time_Evolution}

Again, let us briefly detail how the matrix method is applied:
\begin{itemize}
    \item Use the same perturbing density as eq.~\eqref{eq:Perturbing_Density_Asymptotic}.
    \item Project it onto the basis following eq.~\eqref{eq:BOB_Projection}, and stack the resulting projection in order to build the vector $\ubb$. This stacking of the same vector accounts for the fact that in our case of interest, the perturber is time stationary. For the projections, we used the same basis as in Sec.~\ref{subsubsec:Application_Matrix_FinalState}, but now only $\nmax = 100$ with characteristic radius $\Rb = 3 \, r_{\rs}$. Indeed, integrating the time evolution increases the computational cost of the algorithm. Hence, we had to decrease its spatial resolution.
    \item Independently compute the response matrix of eq.~\eqref{eq:Response_Matrix} from $F_{\ri}$. The action space integral considers all orbits with $0.002 \, r_{\rs} < \rp < \ra < 20 \, r_{\rs}$. The sum over angular Fourier numbers is performed with $n_2 = n_3 = 0$ (spherical symmetry of the background and the perturber) and $|n_1| \leq 2$. The dynamics is integrated for a total time of $10 \, T_{\rs}$, sampled with 600 time steps. We confirmed the convergence of our results w.r.t. changes in these parameters. \cite{Rozier+2022} discuss why such a small number of time steps is sufficient in the framework of LRT.
    \item Apply eq.~\eqref{eq:Linear_response_final} and de-project the response using eq.~\eqref{eq:BOB_Deprojection}.
\end{itemize}
The bottleneck of this procedure is, again, the computation of the response matrix. We use the LiRGHaM code\footnote{\url{https://github.com/simrozier/LiRGHaM}.}, which we complemented with a refined treatment of integral edges as detailed in App.~\ref{app:Edges}. On the fly, we also compute the moment matrices for the mean radial velocity $\bM[v_r]$ and the mean squared velocities $\bM[v_r^2]$ and $\bM[v_{\rt}^2]$, using eq.~\eqref{eq:Moment_Matrix} and the simplifications detailed in App.~\ref{app:Velocity_Moments}. We show the results of the procedure in Sec.~\ref{subsec:Simulations_Time_Evolution}.

\subsection{Comparison of the time evolution with simulations} 
\label{subsec:Simulations_Time_Evolution}

Having developed the framework necessary to model the time evolution of the perturbed system with the matrix method, we now apply this framework to the example disequilibrium system introduced in Sec.~\ref{subsubsec:Model_Near_Equilibrium}. As previously, we compare the results of the matrix calculation against controlled $N$-body models as described in Sec.~\ref{subsubsec:Isolated_NBody_FinalState_setup}. 

\begin{figure}
    \centering
    \includegraphics[width=\columnwidth]{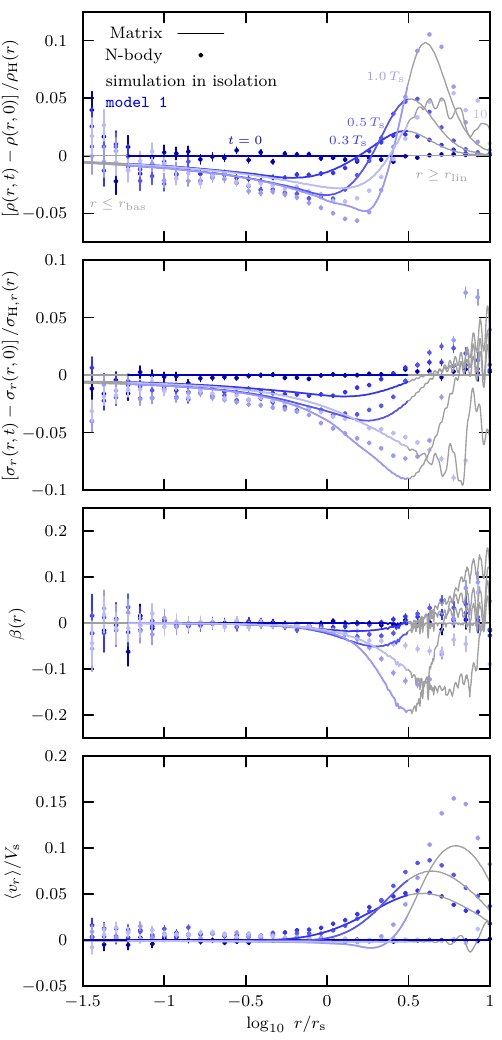}
    \caption{Top panel: Difference between the initial disequilibrium density profile $\rho(t=0,r)$, and the evolving density profile at subsequent times $\rho(t>0,r)$, normalized by the Hernquist density $\rho_\mathrm{H}(r)$. The initial disequilibrium profile corresponds to a truncation in energy parameterised through $\{E_\mathrm{tr}/|E_0|,A/|E_0|\}=\{-0.176,0.077\}$ in Eq.~\ref{eq:truncation_function} (\texttt{model\,1}, see Table~\ref{tab:parameter_overview}). Profiles computed using the matrix method are drawn as solid curves, while results of isolated $N$-body simulation are shown as filled circles. Errorbars indicate the level of Poisson noise expected in the simulation. Second panel: as before, but showing the difference between the radial velocity dispersion profile in the initial disequilibrium state $\sigma_r(t=0,r)$, and the evolved profile $\sigma_r(t>0,r)$. Third panel: anisotropy parameter $\beta = 1 - \sigma^2_{\rt}/(2 \sigma^2_r)$. Bottom panel: mean radial velocity $\langle v_r \rangle$.  }
    \label{fig:time_evolution_comparison}
\end{figure}

Fig.~\ref{fig:time_evolution_comparison} shows the initial ($t=0$) disequilibrium system of \texttt{model~1} ($\{E_\mathrm{tr}/|E_0|,A/|E_0|\}=\{-0.176,0.077\}$ in Eq.~\ref{eq:truncation_function}) in dark blue, as well as in subsequently lighter shades of blue its state after evolving in isolation for a duration of $t=0.3\,T_\mathrm{s}$, $0.5\,T_\mathrm{s}$, $3\,T_\mathrm{s}$ and $10\,T_\mathrm{s}$, where $T_\mathrm{s}$ is defined in Eq.~\ref{Eq.:Ts}. $N$-body snapshots are shown as filled circles, with errorbars indicating the level of Poisson noise. The analytical properties in the initial state, as well as the results of the matrix calculation are shown as solid curves. As before, regions where the matrix method is limited by the resolution of the underlying basis ($r \leq r_\mathrm{bas}$), and regions where non-linear effects may become important ($r \geq r_\mathrm{lin}$) are shown in grey.

The top panel of Fig.~\ref{fig:time_evolution_comparison} shows the difference in density between the initial disequilibrium state  $\rho(t=0,r)$, and the evolved density at subsequent times $\rho(t>0,r)$, normalized by the Hernquist density $\rho_\mathrm{H}(r)$. The time evolution as computed with the matrix method is in agreement with the $N$-body results with per cent-level accuracy for most radii. {The agreement between the $N$-body model and the matrix calculation further improves in the later stages of the evolution, i.e., as the final equilibrium state is approached.} Note the depletion in density in the inner region as a consequence of the relaxation process, occurring very early on in the integration. This feature is further discussed in Sec.~\ref{subsec:Central_Density}.

The second panel shows the difference between the radial velocity dispersion in the initial disequilibrium state $\sigma_r(t=0,r)$, and the evolved states $\sigma_r(t>0,r)$, in units of the Hernquist dispersion $\sigma_\mathrm{H}(r)$. The time evolution as predicted by the matrix method is in good agreement with the $N$-body models in the inner regions. In the outer regions ($r>1.5 r_{\rs}$), while qualitatively similar, the dispersions computed with the matrix method and measured in the $N$-body models differ by several per cent. Also for the case of the velocity dispersion, the agreement between $N$-body model and matrix calculation improves as the final equilibrium state is approached.

The third panel focuses on the time evolution of the anisotropy parameter $\beta = 1 - \sigma^2_{\rt}/(2 \sigma^2_r)$. In the inner regions, the system remains isotropic as it returns to equilibrium. At radii of $1 \leq r_\mathrm{s} \leq 3$, the system develops a mild tangential anisotropy, as seen already in the calculation of the final relaxed state. The matrix calculation and the $N$-body models give qualitatively similar results, but the differences between the results is most pronounced here in the calculation of the anisotropy parameter compared to all other properties studied. 

Finally, the bottom panel of Fig.~\ref{fig:time_evolution_comparison} shows the time evolution of the average radial velocity $\langle v_r \rangle$, in units of the peak circular velocity of the original Hernquist model
\begin{equation}
  V_\mathrm{s} = [G M_\mathrm{tot} / (4\,r_\mathrm{s})]^{1/2}.
  \label{eq:Vs}
\end{equation}
At early times, a large fraction of the system show an outwards motion, as a reaction to the abrupt disappearance of a fraction of the particles. Progressively, the region where $\langle v_r \rangle \approx 0$ grows outwards as the system returns to equilibrium. Indeed, each region of the system relaxes at a rate that follows the local dynamical time. Overall, we see that for radii where the evolution is expected to be linear ($r \leq r_\mathrm{lin}$), the results of the matrix method and the $N$-body models are in excellent agreement.

\section{Discussion} 
\label{sec:Discussion}

\subsection{Comparison with the tidal stripping scenario} 
\label{subsec:Tidal_Stripping}

In recent years, models of the evolution of DM subhalos in a tidal field have challenged the historical picture based on tidal heating. \cite{Amorisco2021} and \cite{Stuecker+2023}, in particular, showed that the remnant subhalo can be accurately modelled by considering that the only effect of the tidal field is to strip a fraction of the system, while the surviving bound remnant is solely evolving due to the loss of these particles, without any direct impact of the tidal field. Here, we reproduce tests of this hypothesis using $N$-body simulations of a Hernquist DM halo in a galactic tidal field, which we compare with results from LRT.

\subsubsection{Numerical setup for the tidal $N$-body simulation}
\label{sec:NbodyTidal}
\textit{$N$-body model.} We generate an $N$-body realization with $N=10^7$ particles of an isotropic Hernquist sphere, as described in Sec.~\ref{subsubsec:Isolated_NBody_FinalState_setup}.

\textit{Host.} We simulate the tidal evolution in an analytical, spherical and static host system with isothermal potential,
\begin{equation}
    \Phi_\mathrm{host}(r) = V_0^2 \ln(r/r_0),
\end{equation}
where $V_0 \approx 33\, V_\mathrm{s}$ (see Eq.~\ref{eq:Vs}) denotes the circular velocity, and $r_0$ is an arbitrary scale radius. 
The Hernquist model is placed at apocentre on an orbit with a pericentric distance of $r_\mathrm{peri} \approx 100\,r_\mathrm{s}$ and a pericentre-to-apocentre ratio of $1:20$. The radial orbital period equals $T_\mathrm{orb} \approx 24\,T_\mathrm{s}$ (see Eq.~\ref{Eq.:Ts}). We chose this radial orbit to ensure that the tidally perturbed model has sufficient time available to return to an (approximate) equilibrium configuration at apocentre. As shown in Fig.~\ref{fig:time_evolution_comparison}, within $\approx 10\,T_\mathrm{s}$, the perturbed model evolved in isolation has returned to a near-equilibrium state ($\langle v_r\rangle \approx 0$) at those radii where the matrix method is applicable -- comparable to the time it takes the tidally perturbed model to return to apocentre after the pericentric passage. 

\textit{Units.} For the sole purpose of illustration we provide an example re-scaling of our model to physical units. For a host circular velocity of $V_\mathrm{c} = 220\,\mathrm{km}\,\mathrm{s}^{-1}$ and a pericentre distance of $r_\mathrm{peri} = 10\,\mathrm{kpc}$, we get $r_\mathrm{apo} = 200\,\mathrm{kpc}$ and $T_\mathrm{orb} \approx 2.3\,\mathrm{Gyr}$. For the Hernquist model, $M_\mathrm{tot} = 4 \times 10^6\,\mathrm{M}_\odot$, $r_\mathrm{s} \approx 0.10\,\mathrm{kpc}$, $V_\mathrm{s}\approx6.6\,\mathrm{km}\,\mathrm{s}^{-1}$ and $T_\mathrm{s} \approx 0.093\,\mathrm{Gyr}$.  

\textit{Particle-mesh code.} We use the same simulation setup as in Sec.~\ref{subsubsec:Isolated_NBody_FinalState_setup}, albeit with a simulation box that encloses the entire orbit. We run the simulation for the duration of a full orbital period within the host potential.

\subsubsection{Initial conditions for the matrix calculation}
\label{sec:MatrixTidal}
We use the results of the $N$-body simulation to inform us about a suiting form of perturbation that may serve as initial conditions to the matrix calculation. For this purpose, we identify the subset of particles located in the final snapshot within a sphere of $5\,r_\mathrm{s}$ around the tidally stripped $N$-body model. This value is chosen purely empirically to approximately define what we consider the virialized remnant system after the pericentric passage, excluding tidal tails and other disequilibrium features. 

To define initial conditions for the matrix calculation, we trace back these selected particles to the simulation's initial conditions. We note that their distribution does not satisfy the hypotheses of eq.~\eqref{eq:Perturbation_Phase_Mixed}, because their distribution has a residual phase dependence and velocity anisotropy. However, we can artificially phase mix and isotropise\footnote{Enforcing initial conditions with isotropic kinematics is a modelling choice we made for the sake of simplicity in this first application of the matrix method to tidal stripping. The matrix method can generally be applied also to anisotropic initial conditions.} it with the following procedure.
First, we phase mix the distribution by computing the differential energy distribution $\rd N / \rd E$ of this subset of particles in the initial Hernquist model. Then, we isotropise it by defining the truncated DF using eq.~\eqref{eq:dNdE}. 
The truncated DF and differential energy distribution are shown using blue filled circles in Fig.~\ref{fig:DF_Hernquist}.  Finally, we fit the truncated DF using the empirical formula Eq.~\ref{eq:truncation_function}, finding the parameters $\{E_\mathrm{tr}/|E_0|,A/|E_0|\} = \{-0.176,0.077\}$ to best match the $N$-body results. 
These initial conditions are identical to the ones used previously as an example in Sec.~\ref{subsec:Simulations_Single_Step} (\texttt{model\,1}, see Tab.~\ref{tab:parameter_overview}); the matrix calculation we compare in the following against the $N$-body run in presence of a tidal field is therefore completely identical to the model shown in Fig.~\ref{fig:final_state_comparison}.

\begin{figure}
    \centering
    \includegraphics[width=\columnwidth]{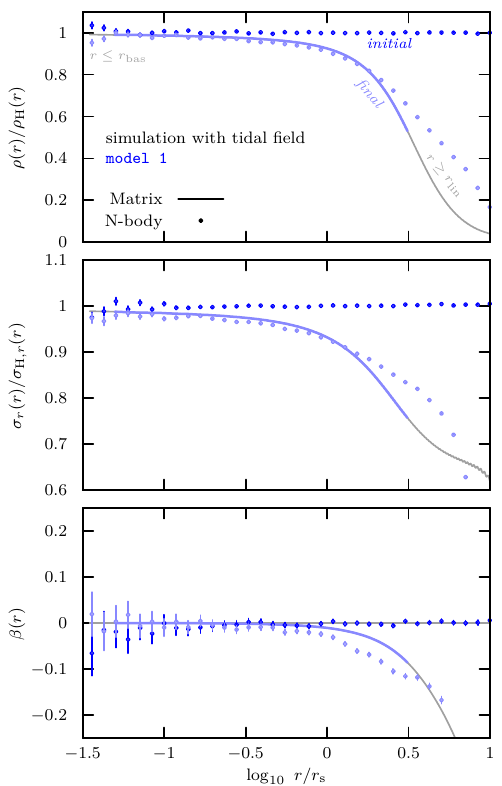}
    \caption{Comparison of \texttt{model\,1} (see Fig.~\ref{fig:final_state_comparison} and Tab.~\ref{tab:parameter_overview}) against an $N$-body realization of a Hernquist profile evolved in presence of a tidal field. Format and units are identical to Fig.~\ref{fig:final_state_comparison}. The ``initial'' $N$-body snapshot shows the model immediately after being injected into the tidal field. The model is injected at the apocentre of an orbit with a pericentre-to-apocentre ratio of 1:20. The ``final'' snapshot shows the $N$-body again at apocentre after a full orbital period ($T_\mathrm{orb}\approx24.2\,T_\mathrm{s}$). In the inner regions, the matrix calculation matches that of the tidally stripped $N$-body model. Deviations are visible in the outer regions, as discussed in the text.}
\label{fig:tidal_field}
\end{figure}

\subsubsection{Comparison of the matrix calculation with a tidal $N$-body simulation}
\label{sec:ComparisonTidal}
We are now able to compare the final equilibrium state as computed using the matrix method with the tidally stripped $N$-body model at apocentre. 
Fig.~\ref{fig:tidal_field} shows density $\rho(r)$, radial velocity dispersion $\sigma_r(r)$ and anisotropy parameter $\beta(r)$ for the evolved $N$-body model (filled circles) and the matrix calculation (solid curve), with units and notations as in Fig.~\ref{fig:final_state_comparison}. For reference, we also show the initial $N$-body Hernquist realization when injected into the host potential. Note that the initial disequilibrium profile used for the matrix calculation is identical to the one shown already in Fig.~\ref{fig:final_state_comparison} and omitted here for clarity. 

In the inner regions ($r \lesssim r_\mathrm{s}$), the simple empirical energy truncation to the initial conditions, evolved to equilibrium using the matrix method, matches the density and radial velocity dispersion as measured in the tidal $N$-body simulation. The results of the $N$-body simulation are consistent with a centrally-isotropic velocity distribution as predicted by the matrix method. The outer anisotropy profile is mildly tangentially biased for both the matrix and the $N$-body model.

Beyond $r_\mathrm{s}$, density and radial velocity dispersion as computed using the matrix method start to deviate from the profiles measured in the $N$-body simulation. Multiple effects may play a role in shaping the outer density profile of the tidally stripped system which are not captured by our simplistic model consisting of an energy truncation and subsequent return to equilibrium. Here, we assume that stripped particles are removed instantaneously from the system, and that the sole response of the system can be captured by it's return to equilibrium in absence of the stripped particles. Our model does not include (1) particles which were stripped and later re-captured, (2) energy injection/tidal heating, which may be important in particular for those particles energetically close to the truncation shown in Fig.~\ref{fig:DF_Hernquist}, (3) any angular momentum dependence when selecting stripped and remaining particles in the initial conditions, (4) non-linear evolution of particles beyond $r > r_\mathrm{lin}$.

\subsection{Depletion of the central density cusp} 
\label{subsec:Central_Density}

In this section, we apply the matrix method to model the tidal evolution of the central density cusp of cold DM subhaloes.
The detailed properties of the central density profile of DM subhaloes are of particular relevance for the study of potential signals of DM self-annihilation and decay, which scale with $\rho^2(r)$ and $\rho(r)$, respectively \citep[see, e.g.][]{Stref2019, Facchinetti2022, Stuecker+2023Annihilation, Delos2023, Lovell2024}. Models of the expected signals need to take into account the evolution of subhalo DM density profiles in the tidal field of the Milky Way. Studies using controlled $N$-body simulations suggest a depletion of the central density \citep[e.g.][]{Hayashi+2003,Penarrubia+2010,Green+2019,Errani+2021}, thereby attenuating any potential emissions from DM self-annihilation. 

The interpretation of $N$-body simulation results however is complicated by the demanding resolution requirements to accurately model the tidal survival of dark substructures. Insufficient (temporal, spatial and particle) resolution of $N$-body simulations is known to heavily impact the tidal evolution of $N$-body subhaloes, affecting both their structure, and their overall abundance \citep[see, e.g.][who coined the term \textit{artificial disruption} for the resolution-driven depletion of dark substructures in simulations]{vandenBoschOgiya2018}. Limited spatial resolution is shown to lead to the formation of a constant-density core even in initially cuspy $N$-body realisations of cold DM haloes \citep{Errani+2020}. Tidal evolution further amplifies this issue: insufficient spatial and/or particle resolution leads to a systematic under-estimation of a subhalo's characteristic density, which in turn renders the subhalo even more vulnerable to the effects of tides \citep{Errani+2021}. Given these resolution challenges, it is no surprise that the depletion of the central density in $N$-body realizations of cuspy DM haloes has been suspected to be a mere numerical artefact, in particular in light of recent analytical approaches predicting virtually unaltered central densities in such haloes \citep{Drakos+2020,Drakos2022,Stuecker+2023}.

\begin{figure}
    \centering
    \includegraphics[width=\columnwidth]{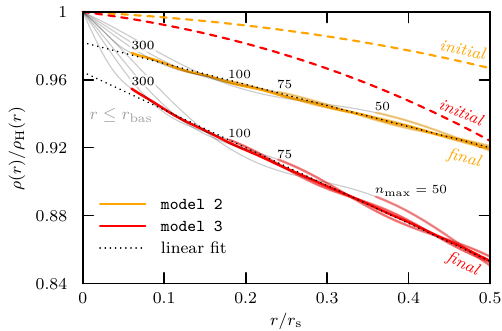}
    \caption{Convergence study. Final density profiles computed using the matrix method for gradually increasing spatial resolution, i.e., number of basis elements $n_\mathrm{max}=50$, $75$, $100$ and $300$. The results obtained for \texttt{model\,2} and \texttt{model\,3} (see Tab.~\ref{tab:parameter_overview}) are shown in orange and red, respectively. 
    The profiles are normalized by the original Hernquist density at each radius. 
    For reference, we also plot the corresponding initial disequilibrium density profiles as dashed curves.
    Radii below the resolution limit $\rbas$ are shown in grey. Within the resolved range of radii, the inner regions of the normalized profiles are well approximated by linear functions. To guide the eye, these are shown as dotted black lines. This convergence test suggests that collisionless relaxation decreases the amplitude of the perturbed Hernquist cusp, even through the initial energy-truncated profile hardly affected the inner regions.} 
\label{fig:central_density}
\end{figure}

Clearly, the matrix method is not subject to the complications arising from discreteness noise, and hence particularly suitable to study the centres of DM subhaloes. The matrix method resolves orbits within the subhalo essentially all the way to the smallest time scales involved. There is, however, a spatial resolution limit intrinsic to the matrix method, arising from the choice of basis. Therefore, we perform convergence tests where we gradually increase the number of basis elements to understand the systematic impact on the central density profile. The results of these tests are show in Fig.~\ref{fig:central_density}. We plot the central density profile in the final equilibrium for \texttt{model\,2} (orange) and \texttt{model\,3} (red) -- see Tab.~\ref{tab:parameter_overview} for parameters. The profiles are normalized by the initial Hernquist density at all radii.
We also plot the initial disequilibrium density profiles as orange and red dashed curves (for \texttt{model\,2} and \texttt{model\,3}, respectively), which both converge to the initial Hernquist density for $r\rightarrow 1$.

We gradually increase the number of basis functions used in the matrix computation, ranging from $n_\mathrm{max}=50$ to $n_\mathrm{max}=300$. The results of this experiment are shown as solid curves in Fig.~\ref{fig:central_density}. In each curve, the resolution limit is taken to be $ \rbas = 3 \, \Rb / \nmax$, this time with $\Rb = 6 \, r_{\rs}$ because it appears to better reconstruct the perturber. Because of the \textit{cored} nature of the basis (every element of the density basis has a finite value at $r=0$), each curve converges to 1 at $r=0$ by construction, i.e., our choice of basis forces the density of the final state at $r=0$ (below the resolution limit $\rbas$) to be identical to the central density of the initial Hernquist model. However, at those radii which we consider well resolved, the density in the final state is always less than in the initial energy-truncated disequilibrium profile. 
For both \texttt{model\,2} and \texttt{model\,3}, the final density in the resolved region is well approximated by linear fits ($\rho(r)/\rho_{\mathrm{H}}(r) \approx 0.98 - 0.12 \, r/r_{\rs}$ for \texttt{model\,2} and $\rho(r)/\rho_{\mathrm{H}}(r) \approx 0.96 - 0.22 \, r/r_{\rs}$ for \texttt{model\,3}), shown in Fig.~\ref{fig:central_density} using back dotted lines.

This convergence test has important implications on how we can interpret the effect of the imposed mass loss. First, the dominant term in the central density is still a cusp with slope $\mathrm{d}\ln \rho /\mathrm{d}\ln r =-1$, so the mass loss did not modify the nature of the central cusp. Second, the amplitude of the cusp has decreased, meaning that the revirialization phase was able to displace an amount of material with divergent central density ($\rho^{\rs} \sim 0.02 \, \rho_{\mathrm{H}}(r) \propto r^{-1}$ and $\rho^{\rs} \sim 0.04 \, \rho_{\mathrm{H}}(r)$ for \texttt{model\,2} and \texttt{model\,3}, respectively), even though the perturber $\rho^{\re}$ has a finite density at the centre.
Coupling this observation with the discussion of Sec.~\ref{subsec:Tidal_Stripping}, our results appear consistent with those obtained with high resolution $N$-body simulations of DM subhalos in smooth tidal fields \citep[see e.g.][]{Green+2019,Errani+2021}: revirialization decreases the cusp's amplitude, however it does not modify its slope\footnote{Note that here, we refer to a progressive tidal stripping process along many pericentric passages in a smooth tidal field. Extreme tidal shocks cannot be modelled with the matrix method as discussed here.}.

Finally, this test case also helps us learn about dynamical studies based on basis function expansions in general. In the matrix method, the quality of the choice of basis is evaluated in light of how well the perturber and the response are reconstructed with as few basis functions as possible. If a basis set is taken to be bi-orthogonal, then usually the central behaviour will be the same for all radial density basis elements. Our problem here is that the central behaviour of the perturber and of the response are very different: the perturber is finite at the centre, while the response diverges as $r^{-1}$. Consequently, there is no single bi-orthogonal basis that is perfectly suited to reconstruct both the perturber and the response. Here, we chose a basis that is finite at the centre, which makes it difficult to detect the response's central divergence; had we chosen a cuspy basis set \citep[e.g.,][]{HernquistOstriker1992,Lilley+2018}, our problem would transform into accurately projecting the perturber. It is likely that the problem persists for tailor made bases, such as ones from \cite{Weinberg1999,Petersen+2022,Lilley+2023,Petersen+2023}.

\section{Conclusions} 
\label{sec:Conclusion}
In the present work, we develop an analytical, linear model for the collisonless relaxation of self-gravitating systems from near equilibrium configurations. 
We apply it to a spherical Hernquist model, which we perturb gently by removing a fraction of the system in the initial conditions. We choose a perturbation in energy, modelled to approximate the type of perturbations typical of weak tidal encounters.  To model the return to equilibrium, we solve the lineralized system of Collisionless Boltzmann -- Poisson equations using a formalism rooted in the \cite{Kalnajs1977} matrix method, which we test against controlled $N$-body simulations.
Below, we summarize our main results.
\begin{itemize}
    \item We show that the final equilibrium state can be computed directly (i.e., in a single computational step) using the matrix method, at very modest computational cost. The equilibrium density profiles agree at sub-percent level with the results of controlled $N$-body simulations. \\
    \item The formalism presented here allows us to compute the full phase space DF, and thereby the velocity anisotropy profile in the final equilibrium state. Comparison against velocity dispersion- and anisotropy profiles measured in controlled $N$-body simulations shows agreement at per cent level. \\
    \item The matrix formalism can also be used to predict the full time evolution from the initial disequilibrium to the final equilibrium, including all intermediate evolutionary stages. The method allows to model the evolution of the density profile, as well as moments of the velocity distribution and derived quantities (such as mean radial velocity, radial velocity dispersion and anisotropy profile). Also here, the evolution predicted by the linear theory is consistent with the results of controlled $N$-body models. \\
    \item We compare the results of the matrix calculation also against $N$-body simulations that include the time-varying tidal field as experienced on a radial orbit in an isothermal host potential. We show that a simple perturbation in energy space, used as initial condition to the matrix method, allows to predict final equilibrium density and velocity dispersion profiles as measured in the $N$-body simulations at radii below the initial Hernquist scale radius. At larger radii, the matrix method and the $N$-body simulations remain qualitatively in agreement and predict a slight tangential bias to the velocity dispersion, though the degree of anisotropy measured in the $N$-body models is larger than the value predicted by our linear theory. 
    As the initial disequilibrium configuration studied here has isotropic kinematics, the tangential anisotropy in the final state is a consequence of the collisionless relaxation alone. \\
    \item The matrix model predicts a decrease in the amplitude of the central density cusp of the tidally perturbed Hernquist sphere, in agreement with the results of previous simulation studies. Crucially, the depletion of the central density is driven by the collisionless relaxation, and not by the initial tidal energy truncation. 
\end{itemize}

The methods developed in this work constitute a computationally highly efficient framework to study the response of gravitational systems to (small) perturbations. We believe the methods may find application in numerous astrophysical contexts. 

In the context of tidal stripping, the methods may be used to efficiently explore the systematics in the response to tidal perturbations for different families of initial distribution functions, such as those of cuspy and cored models with different anisotropy profiles, or rotation. 
Iterative application of the matrix method may allow to approximate the response beyond the linear regime. We will explore these topics in a future contribution.

Furthermore, the method may be applied to study perturbations to the central regions of stellar systems, such as central mass accumulation due to black hole growth in star clusters \citep[e.g.][]{Young1980}, or gas accretion onto galaxies \citep[e.g.]{Sellwood+2005}.
Similarly, the method can be used to model the response to central mass depletion due to stellar feedback, which applies both to the study of cusp-core transformations in dwarf galaxies \citep{Freundlich+2020}, or the (collisionless) response to stellar escape in star clusters. 

Finally, the analytical developments presented here challenge the adiabatic vs. impulsive distinction that is often made in the analysis of the response of stellar systems to (tidal) perturbations. Our results show that in the linear approximation, the time asymptotic equilibrium does not depend on how the perturbation is applied, but solely on its time asymptotic value. These systematics may be fundamental to the emergence of virtually identical tidal evolutionary tracks for cuspy stellar systems subject to impulsive- or adiabatic tidal mass removal.

\section*{Acknowledgments}
The authors would like to thank Anna Lisa Varri and Benoit Famaey for insightful discussions. SR acknowledges support from the Agence Nationale de la Recherche (ANR project GaDaMa ANR-18-CE31-0006) and from the Royal Society (Newton International Fellowship NIF{\textbackslash}R1{\textbackslash}221850).
RE acknowledges support from the European Research Council (ERC) under the European Unions Horizon 2020 research and innovation programme (grant agreement number 834148), and from the National Science Foundation (NSF) grant AST-2206046. Furthermore, the authors gratefully acknowledge the access to computational resources provided by the University of Victoria.

\newpage

\numberwithin{equation}{section} 
\numberwithin{figure}{section}
\appendix

\section{Velocity moments of a spherical, non-rotating system}
\label{app:Velocity_Moments}

\subsection{Simplifying the matrix}

In the case of a spherical potential, the angle Fourier transformed functions $[\mu \psi^{(p)}]_{\bn}(\bJ)$ and $\psi^{(q)}_{\bn}(\bJ)$ appearing in eq.~\eqref{eq:Moment_Matrix} can be simplified, following developments from \cite{Polyachenko1981,FridmanPolyachenko1984(II),TremaineWeinberg1984}. First, we can explicitly develop a basis element as a spherical harmonic times a radial basis function, so that
\begin{equation}
    [\mu \psi^{(p)}]_{\bn}(\bJ) = \!\! \int \!\! \frac{\rd \bT}{(2\pi)^3} \, \mu(\bv) \, U^{\ell^p}_{n^p}(r) Y_{\ell^p}^{m^p}(\theta, \phi) \, \re^{-\ri \bn \cdot \bT}.
\end{equation}
Here, let us highlight that $\theta$ is the polar angle in spherical coordinates, while $\bT = (\theta_1, \theta_2, \theta_3)$ are the angle phase space variables. We can now use the fact that in a spherical potential, a phase space point with coordinates $(r, \theta, \phi, \bv)$ verifies
\begin{align}
\nonumber
    Y_{\ell}^m(\theta, \phi) & = \re^{\ri m \theta_3} \, Y_{\ell}^m(\theta, \phi - \theta_3) \\ & = \re^{\ri m \theta_3} \!\! \sum_{k = -\ell}^{\ell} \!\! \mR_{k m}^{\ell}(\beta) \, y_{\ell}^k \, \re^{\ri k \psi} \, \ri^{m - k},
\end{align}
where $\cos(\beta) = L_z / L$, $\mR$ is Wigner's rotation matrix, $y_{\ell}^k = Y_{\ell}^k(\tfrac{\pi}{2},0)$ and $\psi$ is the angle measured in the orbital plane from the ascending node to the current position. At that point, useful simplifications arise when we assume that $\mu(\bv)$ is only a function of the angle-action variables through the actions $\bJ$ and the radial angle $\theta_1$. In a spherical potential, this is the case if $\mu$ is a function of $v_r$ and $v_{\rt}$, so it applies to all moments of the radial and tangential velocities. Indeed, when we have $\mu(\bJ,\theta_1)$, then
\begin{align}
\nonumber
    [\mu \psi^{(p)}]_{\bn}(\bJ) & = \!\! \sum_{k = -\ell^p}^{\ell^p} \!\! \mR_{k m^p}^{\ell^p}(\beta) \, y_{\ell^p}^k \, \ri^{m^p - k} \\
    \nonumber 
    & \times \!\! \int \!\! \frac{\rd \theta_1}{2\pi} \, \mu(\bJ,\theta_1) U^{\ell^p}_{n^p}(r) \re^{-\ri n_1 \theta_1} \\
    & \times \!\! \int \!\! \frac{\rd \theta_2}{2\pi} \frac{\rd \theta_3}{2\pi} \, \re^{-\ri (n_2 \theta_2 + n_3 \theta_3)} \, \re^{\ri m^p \theta_3} \, \re^{\ri k \psi}.
\end{align}
Now, because $\psi$ does not depend on $\theta_3$, we can carry out the integration w.r.t. $\theta_3$, yielding $\delta_{n_3}^{m^p}$. In addition, we can use the fact that $(\psi - \theta_2)$ is a function of $\theta_1$ only to rewrite
\begin{align}
\nonumber
    [\mu \psi^{(p)}]_{\bn}(\bJ) & = \delta_{n_3}^{m^p} \!\! \sum_{k = -\ell^p}^{\ell^p} \!\! \mR_{k m^p}^{\ell^p}(\beta) \, y_{\ell^p}^k \, \ri^{m^p - k} \\
    \nonumber 
    & \times \!\! \int \!\! \frac{\rd \theta_1}{2\pi} \, \mu(\bJ,\theta_1) U^{\ell^p}_{n^p}(r) \re^{-\ri [n_1 \theta_1 + k(\theta_2 - \psi)]} \\
    & \times \!\! \int \!\! \frac{\rd \theta_2}{2\pi} \, \re^{-\ri (n_2 - k) \theta_2}.
\end{align}
Again, the $\theta_2$ integral gives $\delta_{n_2}^k$, which restricts the sum over $k$ to a single term. Let $\tbn = (n_1, n_2)$ and $g$ be a function of $(\tbJ, \theta_1)$, we can define two operators\footnote{\cite{TremaineWeinberg1984} only define the $W$ operator, because they focus on a choice of $g$ that is merely a function of $r$.}, $W^{\tbn}[g]$ and $Z^{\tbn}[g]$, equal to
\begin{align}
\label{eq:Definition_W_Operator}
    W^{\tbn}[g](\tbJ) & = \frac{1}{2\pi} \!\! \int_{-\pi}^{\pi} \!\!\!\!\! \rd \theta_1 g(\bJ, \theta_1) \cos[n_1 \theta_1 + n_2(\theta_2 - \psi)], \\
    \label{eq:Definition_Z_Operator}
    Z^{\tbn}[g](\tbJ) & = \frac{1}{2\pi} \!\! \int_{-\pi}^{\pi} \!\!\!\!\! \rd \theta_1 g(\bJ, \theta_1) \sin[n_1 \theta_1 + n_2(\theta_2 - \psi)].
\end{align}
With these definitions, we have
\begin{align}
\nonumber
    [\mu \psi^{(p)}]_{\bn}\!(\bJ) & \!=\! \delta_{n_3}^{m^p} \mR_{n_2 m^p}^{\ell^p}(\beta) \, y_{\ell^p}^{n_2} \, \ri^{m^p - n_2} \\ 
    & \times \big(W^{\tbn}[\mu U_{n_p}^{\ell^p}](\tbJ) - \ri \, Z^{\tbn}[\mu U_{n^p}^{\ell^p}](\tbJ) \big).
\end{align}

From eqs.~\eqref{eq:Definition_W_Operator} and~\eqref{eq:Definition_Z_Operator}, we see that because $\theta_2 - \psi$ is odd in $\theta_1$, then $W^{\tbn}[g] = 0$ (resp. $Z^{\tbn}[g] = 0$) for any $g$ that is odd (resp. even) in $\theta_1$. Decomposing $g = g_+ + g_-$ in even and odd parts, we therefore have
\begin{align}
\label{eq:W_Op}
    W^{\tbn}[g](\tbJ) & = \frac{1}{\pi} \!\! \int_{0}^{\pi} \!\!\!\!\! \rd \theta_1 g_+(\bJ, \theta_1) \cos[n_1 \theta_1 + n_2(\theta_2 - \psi)], \\
    \label{eq:Z_Op}
    Z^{\tbn}[g](\tbJ) & = \frac{1}{\pi} \!\! \int_{0}^{\pi} \!\!\!\!\! \rd \theta_1 g_-(\bJ, \theta_1) \sin[n_1 \theta_1 + n_2(\theta_2 - \psi)].
\end{align}
We also note that they verify the symmetry relations
\begin{equation}
    W^{-\tbn} = W^{\tbn} \quad ; \quad Z^{-\tbn} = -Z^{\tbn}.
\end{equation}
Finally, since $r$ is an even function of $\theta_1$, we have
\begin{align}
\nonumber
    [\mu \psi^{(p)}]_{\bn}\!(\bJ) & \!=\! \delta_{n_3}^{m^p} \mR_{n_2 m^p}^{\ell^p}(\beta) \, y_{\ell^p}^{n_2} \, \ri^{m^p - n_2} \\ 
    & \times \big(W^{\tbn}[\mu_+ U_{n^p}^{\ell^p}](\tbJ) - \ri \, Z^{\tbn}[\mu_- U_{n^p}^{\ell^p}](\tbJ) \big).
    \label{eq:Fourier_Transformed_Basis}
\end{align}
We conveniently recover eq.~(A4) of \cite{Rozier2019} by taking $\mu(\bv) = 1$. 

Further simplifications are obtained when we assume that $F_{\ri}$ is a function of the actions only through $\tbJ = (J_r, L)$, implying that the system is non-rotating. Then, the only $L_z$ dependence in the integrand of eq.~\eqref{eq:Moment_Matrix} resides in the $\mR_{n_2 m^p}^{\ell^p}(\beta) \mR_{n_2 m^p}^{\ell^q}(\beta)$ terms from the angular Fourier transforms. Owing to the orthogonality relations of the Wigner rotation matrices, the $L_z$ integral can then be simplified. Thus, the response matrix can be rewritten
\begin{equation}
    \bM[\mu]_{pq}(t) = \delta_{m^p}^{m^q} \delta_{\ell^p}^{\ell^q} \sum_{\tbn} C_{\ell^p}^{n_2} \, P[\mu]_{\ell^p n^p n^q}^{\tbn}(t),
\label{eq:Moment_Matrix_Final}
\end{equation}
where the coefficients $C$ are given by
\begin{equation}
    C_{\ell}^n = 2 (2\pi)^3 \frac{ |y_{\ell}^{n}|^2}{2 \ell + 1},
\end{equation}
and the functions $P$ by
\begin{align}
\nonumber
    P[\mu]_{\ell^p n^p n^q}^{\tbn}(t) & = \!\! \int \!\! \rd \tbJ \, \re^{-\ri \tbn \cdot \tbO t} \, L \, \tbn \cdot \frac{\partial F_{\ri}}{\partial \tbJ} \, W^{\tbn}[U_{n^q}^{\ell^p}] \\
    & \times \big(Z^{\tbn}[\mu_-U_{n^p}^{\ell^p}] - \ri \, W^{\tbn}[\mu_+U_{n^p}^{\ell^p}]\big).
    \label{eq:Functions_P}
\end{align}
A final computational speedup is achieved by noticing that 
\begin{equation}
    C_{\ell}^{-n} = C_{\ell}^{n} \quad ; \quad P^{-\tbn} = (P^{\tbn})^*,
\end{equation}
so that the sum in eq.~\eqref{eq:Moment_Matrix_Final} can be easily reduced to half of the possible $\tbn$ pairs.

\subsection{Numerical implementation}
\label{subsec:Numerical_Details}

Here, we summarize the numerical implementation of the response matrix computation, referring the reader to \cite{Rozier2019} for more details. 

The numerical effort required to compute the response matrix of eq.~\eqref{eq:Moment_Matrix_Final} is essentially contained in the computation of the functions $P$ of eq.~\eqref{eq:Functions_P}. Three specific difficulties must be highlighted: (i) performing the action-space integration; (ii) computing the orbital characteristics (actions, frequencies) of each orbit evaluated; (iii) computing the angle-space integrals of eqs.~\eqref{eq:W_Op} and~\eqref{eq:Z_Op}.

\textit{Action-space integration.} In order to simplify the computation of the orbital elements, we change variables of the action-space integrals to the orbital peri- and apocentre $(\rp, \ra)$. Because of the structure of the integrand, we perform an extra change of variables into a new set $(u,v)$ that is denser in two regions: when $\ra \ll r_{\rs}$ (central region) and when $\ra - \rp \ll \rp$ (nearly circular orbits). For numerical efficiency, $\rp(u,v)$ and $\ra(u,v)$ are explicit functions that are easy to compute and derive. Hence, if we formally write eq.~\eqref{eq:Functions_P} as
\begin{equation}
    P = \!\! \int \!\! \rd \tbJ \, p(\tbJ),
\end{equation}
then we re-express the integral so that
\begin{equation}
    P = \!\! \int \!\! \rd u \rd v \, \bigg|\frac{\partial \tbJ}{\partial \rp,\ra}\bigg| \, \bigg|\frac{\partial \rp,\ra}{\partial u,v}\bigg| \, p(\tbJ(\rp(u,v),\ra(u,v))),
\end{equation}
where we included the respective Jacobian determinants of the changes of variables. 

Then, the integration on the $(u,v)$ plane is performed via a tailor-made method that is adapted to the specific structure of the integrand. In more detail, the integral can be rewritten as
\begin{equation}
    P = \!\! \int \!\! \rd u \rd v \, g(u,v) \, \re^{\ri h(u,v)}, 
\end{equation}
where both $g$ and $h$ are slowly varying functions of their arguments. Therefore, we divide the $(u,v)$ plane in small squares of side $\Delta u$ centered around grid points $(u_i, v_i)$, and within each square we perform a bi-linear expansion of $g$ and $h$ around $(u_i, v_i)$. The resulting integrand is
\begin{equation}
    \int_{- \tfrac{\Delta u}{2}}^{\tfrac{\Delta u}{2}} \!\!\!\! \rd x \!\! \int_{- \tfrac{\Delta u}{2}}^{\tfrac{\Delta u}{2}} \!\!\!\! \rd y \, \big( g + g_u x + g_v y \big) \, \re^{\ri (h + h_u x + h_v y)},
\end{equation}
where the subscripts $u$ and $v$ denote respective derivatives w.r.t. $u$ and $v$, and all functions are evaluated at the point $(u_i, v_i)$. This integral has an analytical expression as a function of $g(u_i,v_i),g_u(u_i,v_i),g_v(u_i,v_i),h(u_i,v_i),h_u(u_i,v_i),h_v(u_i,v_i)$. Once these six evaluations are made for each grid point, we can simply add the contribution from each square to compute the total action-space integral. All our results are checked for convergence w.r.t. the grid step size, $\Delta u$. 

\textit{Orbital characteristics.} In order to compute the values of $g, h$ and their derivatives on the $(u,v)$ grid, one needs to specify how the orbital characteristics (actions, frequencies and their derivatives) are computed. At each grid point, we straightforwardly obtain the values of $(\rp(u_i,v_i), \ra(u_i,v_i))$, from which all other functions can be derived. Specifically, we obtain the orbital energy and angular momentum through 
\begin{equation}
    E \!=\! \frac{\ra^2 \psi_0(\ra) - \rp^2 \psi_0(\rp)}{\ra^2 - \rp^2}; L \!=\! \sqrt{\frac{2 (\psi_0(\ra) - \psi_0(\rp))}{\rp^{-2} - \ra^{-2}}},
\end{equation}
the radial action through
\begin{equation}
    J_r = \frac{1}{\pi} \!\! \int_{\rp}^{\ra} \!\! \rd r \, \sqrt{2(E - \psi_0(r)) - L^2 / r^2}, 
\end{equation}
and the frequencies through
\begin{align}
    \Omega_1 & = \bigg[ \frac{1}{\pi} \!\! \int_{\rp}^{\ra} \!\! \rd r \frac{1}{\sqrt{2(E - \psi_0(r)) - L^2 / r^2}} \bigg]^{-1}, \\
    \Omega_2 & = \frac{\Omega_1}{\pi} \!\! \int_{\rp}^{\ra} \!\! \rd r \frac{L / r^2}{\sqrt{2(E - \psi_0(r)) - L^2 / r^2}}.
\end{align}
To perform the radial integrals, we first change variables to an orbital anomaly $x$, i.e. any integral becomes
\begin{equation}
    Y(\rp, \ra) \!=\! \!\!\int_{\rp}^{\ra}\!\!\!\! \rd r \, y(r,\rp,\ra) \!=\! \!\! \int_{-1}^1 \!\!\! \rd x \, \frac{\rd r}{\rd x} \, y(r(x), \rp, \ra).
\end{equation}
We tune the change of variables so that all integrals run from $-1$ to $1$ and any integrable divergence is cured. Then, any derivative of $Y$ w.r.t. $\rp$ or $\ra$ can be straightforwardly transported under the integral sign, so that they can also be written in that integral form. Then, we perform the integration through a Gauss-Kronrod method. 

\textit{Angle-space integrals.} The angle-space integral requires particular care, because it is the bottleneck of the whole matrix evaluation. Let us focus on eq.~\eqref{eq:W_Op}. Again, the most direct way of computing the integral is to change orbital variables from radial angle to radius, because we do not have an explicit expression for the azimuthal anomaly $\theta_2 - \psi$ as a function of $\theta_1$. Hence, we rewrite the function $W$ as
\begin{equation}
    W = \!\! \int_0^{\pi} \!\! \rd \theta_1 \, w(\theta_1,\theta_2) =  \!\! \int_{\rp}^{\ra} \!\!\!\! \rd r \, \frac{\rd \theta_1}{\rd r} \, w(\theta_1(r),\theta_2(r)).
\end{equation}
Here, the angles have an explicit integral expression as a function of $r$, as
\begin{align}
    \theta_1(r) & = \!\! \int_{\rp}^{r} \!\! \rd r^{\prime} \frac{\Omega_1}{\sqrt{2(E - \psi_0(r^{\prime})) - L^2 / r^{\prime 2}}}, \\
    (\theta_2 - \psi)(r) & = \!\! \int_{\rp}^{r} \!\! \rd r^{\prime} \frac{\Omega_2 - L / r^{\prime 2}}{\sqrt{2(E - \psi_0(r^{\prime})) - L^2 / r^{\prime 2}}}.
\end{align}
Because the integration boundaries of $\theta_1, \theta_2$ match the integration variable of $W$, the integral for $W$ can be computed as a single 1D integral instead of a 2D nested integral, via a method detailed in \cite{Rozier2019}. Similarly to the orbital characteristics, the integrals over $r$ are performed via a change of variables with an orbital anomaly $x$, which also helps computing the derivatives of $W$ w.r.t. $\rp, \ra$.

\subsection{Numerical complexity}
\label{subsec:Complexity}

We quantify here the numerical cost of the matrix method. We first focus on one of the runs described in Section~\ref{subsubsec:Isolated_NBody_FinalState_results}, and we give scaling relations that help estimating the complexity of other runs in the paper.

Computing the response matrix for \texttt{model\,1} in Fig.~\ref{fig:final_state_comparison} involved $2\times10^{15}$ floating-point operations, and was one of the most involved runs of the paper, as it required both a sufficient dynamic range in radii and resolution in the centre. The radial resolution can be modified by tuning $\nmax$, the numerical complexity changing as $\nmax^4$. This dependence is due to the fact that $\nmax$ impacts both the size of the response matrix (as $\nmax^2$) and the resolution of the action space integration grid (as $\nmax^2$). 

In the runs of Section~\ref{subsec:Simulations_Time_Evolution}, the number of operations is impacted in an affine way by the number of time steps. By lowering the radial resolution (through $\nmax$) and the maximum order of the angular Fourier number $n_1$, we were able to keep the number of floating-point operations at around $2\times10^{15}$ as well.

\section{Edge integrals}
\label{app:Edges}

\subsection{Definition}

In our case of interest, a significant contribution to the response matrix can come from integrating eq.~\eqref{eq:Response_Matrix} at the edges of the integration domain\footnote{The exact same methods are applied to compute the time asymptotic matrix of eq.~\eqref{eq:Response_Matrix_Asymptotic}, or the moment response matrix of eq.~\eqref{eq:Moment_Matrix}.}. As shown in \cite{JalaliHunter2005} \citep[see also][]{PolyachenkoShukhman2015}, the edge integrals can be considered by replacing $F_{\ri}$ in the integrand with $F_{\ri} \times \mathcal{I}(\bJ)$, where $\mathcal{I}$ is the indicator function of the integration domain (yielding 1 inside and 0 outside). Consequently, a term has to be added to $\bM_{pq}(t)$, equal to 
\begin{equation}
    - \ri (2 \pi)^3 \! \sum_{\bn} \!\!\int\!\! \rd \bJ \, \re^{- \ri \, \bn \cdot \bO \, t} \, F_{\ri}(\bJ) \, \bn \!\cdot\! \frac{\partial \mathcal{I}}{\partial \bJ} \psi_{\bn}^{(p)*\!}(\bJ) \psi_{\bn}^{(q)\!}(\bJ).
\end{equation}
In practice, as explained in Section~\ref{subsec:Numerical_Details}, the action space integral reduces to a 2D integral over $\tbJ = (J_r, L)$, and it is performed via successive changes of variables, from $J_r, L$ through $E,L$, then $\rp, \ra$ and eventually $u,v$. Hence, the function $\mathcal{I}$ is defined in terms of these $u,v$ variables. Simple but tedious algebra helps us compute the derivatives of $\mathcal{I}$ w.r.t. the actions from its derivatives w.r.t. $u,v$, yielding
\begin{align}
\nonumber
    \tbn \cdot \frac{\partial \mathcal{I}}{\partial \tbJ} = \bigg| \!\!\!&\,\,\, \frac{\partial E,L}{\partial \rp, \ra} \bigg|^{-1} \, \bigg| \frac{\partial \rp, \ra}{\partial u,v} \bigg|^{-1} \\
    \nonumber 
    \times \Bigg(\! \tbn \cdot \tbO & \bigg[\! \bigg(\! \frac{\partial \mathcal{I}}{\partial u} \frac{\rd (\ra\!-\!\rp)}{\rd v} \!-\! \frac{\partial \mathcal{I}}{\partial v} \frac{\rd \rp}{\rd u} \!\bigg)\! \frac{\partial L}{\partial \ra} \!-\! \frac{\partial \mathcal{I}}{\partial v} \frac{\rd \rp}{\rd u} \frac{\partial L}{\partial \rp} \!\bigg] \\
     - n_2 & \bigg[\! \bigg(\! \frac{\partial \mathcal{I}}{\partial u} \frac{\rd (\ra\!-\!\rp)}{\rd v} \!-\! \frac{\partial \mathcal{I}}{\partial v} \frac{\rd \rp}{\rd u} \!\bigg)\! \frac{\partial E}{\partial \ra} \!-\! \frac{\partial \mathcal{I}}{\partial v} \frac{\rd \rp}{\rd u} \frac{\partial E}{\partial \rp} \!\bigg] \!\Bigg)\!.
\end{align}
Here, the two first terms in the r.h.s. are the Jacobian determinants of the transformations $(\rp,\ra)\rightarrow (E,L)$ and $(u,v) \rightarrow (\rp,\ra)$, and we made use of the fact that in our specific change of variables, $\ra(u,v) = \rp(u) + (\ra - \rp)(v)$, so that $\partial \rp / \partial v = 0$ and $\partial \ra / \partial u = \partial \rp / \partial u$. It therefore remains to specify the derivatives of $\mathcal{I}$ w.r.t. $u,v$ at the edges of the integration domain.

Four cases arise, depending on the specific edge considered: (i) $u = \umin(v)$, where locally $\mI = \Theta(u - \umin(v))$, (ii) $u = \umax(v)$, where $\mI = \Theta(\umax(v) - u)$, (iii) $v = \vmin(u)$, where $\mI = \Theta(v - \vmin(u))$, and (iv) $v = \vmax(u)$, where $\mI = \Theta(\vmax(u) - v)$. More details on the way these edges relate to the input parameters can be found in \cite{Rozier2019}, we summarize here their main characteristics. 
\begin{itemize}
    \item[(i)] $u = \umin(v)$ is reached when $\rp(u)$ is maximal at fixed $v$, corresponding to when $\ra$ is maximal: $\ra(\umin(v),v) = \rmax$. So $\umin(v) = \rp^{-1}(\rmax - (\ra - \rp)(v))$. As a consequence, 
    \begin{equation}
        \frac{\rd \umin}{\rd v} = - \frac{\tfrac{\rd (\ra-\rp)}{\rd v}}{\tfrac{\rd \rp}{\rd u}(\umin)}.
    \end{equation}
    Hence, on that edge, we have
    \begin{align}
    \nonumber
        \frac{\partial \mI}{\partial u} & = \delta(u - \umin(v)) ; \\  \frac{\partial \mI}{\partial v} & = \frac{\tfrac{\rd (\ra-\rp)}{\rd v}}{\tfrac{\rd \rp}{\rd u}(\umin)} \delta(u - \umin(v)).
    \end{align}
    \item[(ii)] $u = \umax(v)$ is reached when $\rp(u) = \rmin$. Hence
    \begin{equation}
        \frac{\rd \umax}{\rd v} = 0,
    \end{equation}
    and we have
    \begin{align}
    \nonumber
        \frac{\partial \mI}{\partial u} & = - \delta(\umax(v) - u) ; \\  \frac{\partial \mI}{\partial v} & = 0.
    \end{align}
    \item[(iii)] $v = \vmin(u)$ is reached when $\ra(u,\vmin(u)) = \rmax$. So $\vmin(u) = (\ra - \rp)^{-1}(\rmax - \rp(u))$. As a consequence,
    \begin{equation}
        \frac{\rd \vmin}{\rd u} = - \frac{\tfrac{\rd \rp}{\rd u}}{\tfrac{\rd (\ra - \rp)}{\rd v}(\vmin)}.
    \end{equation}
    Hence, on that edge, we have
    \begin{align}
    \nonumber
        \frac{\partial \mI}{\partial u} & = \frac{\tfrac{\rd \rp}{\rd u}}{\tfrac{\rd (\ra - \rp)}{\rd v}(\vmin)} \delta(v - \vmin(u)) ; \\  \frac{\partial \mI}{\partial v} & = \delta(v - \vmin(u)).
    \end{align}
    \item[(iv)] Finally, $\vmax$ is reached for $(\ra - \rp)(v) = \rmin$, so that 
    \begin{align}
    \nonumber
        \frac{\partial \mI}{\partial u} & = 0 ; \\  \frac{\partial \mI}{\partial v} & = - \delta(\vmax(u)-v).
    \end{align}
\end{itemize}

\subsection{Numerical implementation}

On each of the edges of the $(u,v)$ integration domain, we have to perform a 1D integral of the form
\begin{equation}
    \mJ = \!\!\int\!\! \rd v \, g(\umin(v),v) \, \re^{\ri \, h(\umin(v),v)}.
\end{equation}
In the same way as our method for computing 2D integrals with a fast trigonometric oscillation, we cut the integration domain in small segments of width $\Delta v$ centered on successive values $v_i$. Then, we expand the slowly varying functions $g, h$ to first order in $v$ around $v_i$, so that we approximate
\begin{align}
\nonumber
    \mJ & = \sum_{v_i} \!\!\int_{- \tfrac{\Delta v}{2}}^{\tfrac{\Delta v}{2}}\!\!\! \rd y \, \big[ g + \big(g_u \tfrac{\rd \umin}{\rd v} + g_v\big) y \big] \, \re^{\ri \big[h + \big(h_u \tfrac{\rd \umin}{\rd v} + h_v\big) y\big]} \\
    & =  \sum_{v_i} g \, \re^{\ri h} \, \aleph\bigg(\frac{g_u \tfrac{\rd \umin}{\rd v} + g_v}{g}, h_u \tfrac{\rd \umin}{\rd v} + h_v, \Delta v \bigg),
\end{align}
where all functions are evaluated at $(\umin(v_i),v_i)$ and $\aleph$ is given by
\begin{equation}
    \aleph(a,b,\Delta v) = \!\! \int_{- \tfrac{\Delta v}{2}}^{\tfrac{\Delta v}{2}} \!\! \rd v \, (1 + a v) \, \re^{\ri b v}.
\end{equation}
This can be further simplified by considering the integral with normalized bounds, i.e.
\begin{equation}
    \aleph(a,b,\Delta v) = \Delta v \, \aleph_{\rD}(a \Delta v, b \Delta v),
\end{equation}
where 
\begin{equation}
    \aleph_{\rD}(\alpha, \beta) = \!\! \int_{-1/2}^{1/2} \!\! \rd x \, (1 + \alpha x) \, \re^{\ri \beta x}.
\end{equation}
This integral has a straightforward analytical expression in terms of trigonometric functions of $\alpha$ and $\beta$. The same steps can be followed to adapt this development to the other edges of the integration domain.

\footnotesize
\bibliography{matrix}{}
\bibliographystyle{aasjournal}

\end{document}